\documentclass[aps,prd,twocolumn,citeautoscript,superscriptaddress,nofootinbib]{revtex4-1}  
\usepackage[colorlinks=true]{hyperref}  
\hypersetup{
    bookmarks=true,         
    unicode=false,          
    pdftoolbar=true,        
    pdfmenubar=true,        
    pdffitwindow=false,     
    pdfstartview={FitH},    
    pdftitle={Topological order in the pseudogap metal},    
    pdfauthor={Mathias Scheurer, Shubhayu Chatterjee, Michel Ferrero, Antoine Georges, Subir Sachdev, Wei Wu},     
    pdfsubject={},   
    pdfcreator={},   
    pdfproducer={}, 
    pdfkeywords={} {} {}, 
    pdfnewwindow=true,      
    colorlinks=true,       
    linkcolor=magenta, 
    citecolor=blue,        
    filecolor=magenta,      
    urlcolor=blue           
}

\usepackage{fancyhdr, amssymb, bm, amsmath, graphicx, tikz}
\usepackage{isomath}

\usepackage{amsmath}
\usepackage{amssymb}
\usepackage{bbm}
\usepackage{braket}
\usepackage{xcolor}

\newcommand{\beq}{\begin{equation}}
\newcommand{\eeq}{\end{equation}}
\def\bea{\begin{eqnarray}}
\def\eea{\end{eqnarray}}

\newcommand{\equref}[1]{Eq.~(\ref{#1})}
\newcommand{\equsref}[2]{Eqs.~(\ref{#1}) and (\ref{#2})}

\newcommand{\figref}[1]{Fig.~\ref{#1}}
\newcommand{\refcite}[1]{\cite{#1}}

\newcommand{\appref}[1]{{Appendix~\ref{App#1}}}

\newcommand{\pdagger}{{\phantom{\dagger}}}
\newcommand{\diff}{\mathrm{d}}

\renewcommand{\approx}{\simeq}

\renewcommand{\vec}[1]{\boldsymbol{#1}}
\newcommand{\eqtx}[1]{{\color{black}#1}} 

\newcommand{\ie}{\textit{i.e.}}

\newcommand{\ChDisp}{\rho}
\newcommand{\retarded}{r}
\newcommand{\change}[1]{{\color{black} #1}}

\begin{document}

\title{Topological order in the pseudogap metal}

\author{Mathias S.~Scheurer}
\affiliation{Department of Physics, Harvard University, Cambridge MA 02138, USA}

\author{Shubhayu Chatterjee}
\affiliation{Department of Physics, Harvard University, Cambridge MA 02138, USA}

\author{Wei Wu}
\affiliation{Centre de Physique Th\'eorique, \'Ecole Polytechnique, CNRS, Universit\'e Paris-Saclay, 91128 Palaiseau, France}
\affiliation{Coll\`ege de France, 11 place Marcelin Berthelot, 75005 Paris, France}

\author{Michel Ferrero}
\affiliation{Centre de Physique Th\'eorique, \'Ecole Polytechnique, CNRS, Universit\'e Paris-Saclay, 91128 Palaiseau, France}
\affiliation{Coll\`ege de France, 11 place Marcelin Berthelot, 75005 Paris, France}

\author{Antoine Georges}
\affiliation{Centre de Physique Th\'eorique, \'Ecole Polytechnique, CNRS, Universit\'e Paris-Saclay, 91128 Palaiseau, France}
\affiliation{Center for Computational Quantum Physics, Flatiron Institute,  
162 Fifth Avenue, New York, NY 10010, USA} 
\affiliation{Institut de Physique, Coll\`ege de France, 11 place Marcelin Berthelot, 75005 Paris, France}
\affiliation{DQMP, Universit\'e de Gen\`eve, 24 quai Ernest Ansermet, CH-1211 Gen\`eve, Suisse}

\author{Subir Sachdev}
\affiliation{Department of Physics, Harvard University, Cambridge MA 02138, USA}
\affiliation{Perimeter Institute for Theoretical Physics, Waterloo, Ontario, Canada N2L 2Y5}
\affiliation{Department of Physics, Stanford University, Stanford CA 94305, USA}

\begin{abstract}
We compute the electronic Green's function of the topologically ordered Higgs phase of a SU(2) gauge theory of fluctuating antiferromagnetism on the square lattice. The results are compared with cluster extensions of dynamical mean field theory, and quantum Monte Carlo calculations, on the pseudogap phase of the strongly interacting hole-doped Hubbard model. Good agreement is found in the momentum, frequency, hopping, and doping dependencies of the spectral function and electronic self-energy. We show that lines of (approximate) zeros of the zero-frequency electronic Green's function are signs of the underlying topological order of the gauge theory, and describe how these lines of zeros appear in our theory of the Hubbard model. We also derive a modified, non-perturbative version of the Luttinger theorem that holds in the Higgs phase.
 \end{abstract}

\maketitle

\section{Introduction}
\label{sec:intro}

The pseudogap metal is a novel state of electronic matter found in the hole-doped, cuprate high temperature superconductors \cite{Keimer15}.  
It exhibits clear evidence of electrical transport with the temperature and frequency dependence of a conventional metal obeying Fermi liquid theory \cite{Marel13,MG14}.
However, a long-standing mystery in the study of the cuprates is that photoemission experiments do not show the `large' Fermi surface that is expected
from the Luttinger theorem of Fermi liquid theory \cite{LuttingerWard}. (Broken square lattice translational symmetry can allow
`small' Fermi surfaces, but
there is no sign of it over a wide range of temperature and doping over which the pseudogap state is present \cite{Keimer15,LTCP15},
and we will not discuss states with broken symmetry here.)
There are non-perturbative arguments \cite{MO00,FFL,TSMVSS04,APAV04,SSreview18} that deviations from the
Luttinger volume are only possible in quantum states with topological order. But independent evidence for the presence of topological order in the
pseudogap has so far been lacking.

In this paper we employ a SU(2) gauge theory of fluctuating antiferromagnetism (AF)
in metals \cite{SS09,OurPreprint} to describe the pseudogap metal. Such a gauge theory describes fluctuations in the {\it orientation \/} of the AF order, while preserving a local, non-zero magnitude. \change{The emergent gauge fields of the theory indicate the long-range quantum entanglement of the topologically ordered phase.}
An alternative, semiclassical treatment of fluctuations 
of the AF order parameter has been used to describe the electron-doped cuprates \cite{Tremblay04b}, but this remains valid
at low temperatures ($T$) only if the AF correlation length $\xi_{\text{AF}}$ diverges as $T \rightarrow 0$.
We are interested in the case where $\xi_{\text{AF}}$ remains finite at $T=0$, and then a gauge theory formulation is required to keep proper track of the fermionic degrees of freedom in the background of the fluctuating AF order.
Such a gauge theory can formally be derived from a lattice Hubbard model, as we will outline in the next section.
The SU(2) gauge theory yields a pseudogap metal with only `small' Fermi surfaces when the gauge group is `Higgsed' down
to a smaller group. We will describe examples of Higgsing down to U(1) and $\mathbb{Z}_2$, and these will yield  metallic
states with U(1) and $\mathbb{Z}_2$ topological order. See \appref{A} for a definition of topological order in gapless systems; for the U(1) case we primarily consider, the topological order is associated \cite{NRSS90,Hermele04} with the suppression of `hedgehog' defects in the spacetime configuration of the fluctuating AF order. 

We will present a mean-field computation of the electronic Green's function across the entire Brillouin zone in the U(1) Higgs phase of the SU(2) gauge theory.
Such results allow for a direct comparison with numerical computations on the Hubbard model. One of our main results will be that for a reasonable range of parameters in the SU(2) gauge theory,
{\it both} the real and imaginary parts of the electron Green's function of the 
gauge theory with topological order closely resemble those
obtained from the dynamical cluster approximation (DCA), a cluster extension of dynamical mean field theory (DMFT) \cite{georges1996review,tremblay2006pseudogap,kotliar_CDMFT_prl_2001,maier2005review}.
While DCA allows us to study the regime of strong correlations down to low temperature, it has limited momentum-space resolution. For this reason, we have also performed determinant quantum Monte Carlo (DQMC) calculations and find self-energies that, in the numerically accessible temperature range, agree well with the gauge theory computations. Additional results on the comparison between the SU(2) gauge theory and the DCA and DQMC computations, as a function of doping and second-neighbor hopping, appear in a 
companion paper \cite{Georges17a}.

In several discussions in the literature \cite{ACDF,ET02,ID03,KRT06,YRZ,KotliarZeros,DPK12,berthod2006,sakai2009evolution}, violations of the Luttinger theorem have been linked to the presence of lines of zeros (in two spatial dimensions)  
in the electron Green's function on a ``Luttinger surface''. The conventional perturbative proof of the Luttinger
theorem yields an additional contribution to the volume enclosed by the Fermi surface when the electron Green's function has lines of zeros:
it was therefore argued that metallic states with small Fermi surfaces are permitted, even in the context of this perturbative proof of the Luttinger theorem. But this argument appears problematic because the real part of the 
Green's function (and hence the positions of the zeros) can be changed by modifying the spectral density at 
high frequencies, and so it would appear that high energy excitations have an undue influence on the low
energy theory. 
Indeed, specific computations of higher
order corrections \cite{ACDF} do find that lines of zeros in metallic Green's \change{functions} do not contribute to the volume enclosed by the Fermi surface.

We maintain that violations of the Luttinger theorem in metals cannot appear in states that are perturbatively accessible from the free electron state, but are only possible in 
non-perturbative metallic states with topological order \cite{MO00,FFL,TSMVSS04,APAV04}.
While the SU(2) gauge theory is shown to violate the conventional Luttinger theorem, we derive a modified sum rule on the Fermi surfaces of the metallic states with topological order.

We also find that the non-zero temperature SU(2) gauge theory has
lines of {\it approximate\/} zeros of the electron Green's function in a suitable regime. These lines are remnants of lines of zeros
in the mean-field Green's functions of {\it fractionalized\/} particles (`chargons') in the theory. So we claim
that the approximate zeros can be
interpreted as heralds of the underlying topological order. 
Moreover, the good agreement between the SU(2) gauge theory and the numerical computations in DCA and DQMC,
noted above, appears in the regime where the approximate lines of zeros are present. 
Taken together, we reach
one of our main conclusions: there is evidence for topological order in the DCA and DQMC studies 
of the pseudogap state of the Hubbard model.


\section{SU(2) gauge theory of the Hubbard model}
\label{sec:su2}
We are interested here in the Hubbard model with Hamiltonian
\begin{equation}
\hat{H}_U = -\sum_{i,j} t_{ij} \hat{c}^\dagger_{i\alpha} \hat{c}^\pdagger_{j\alpha} - \mu \sum_i \hat{c}^\dagger_{i\alpha} \hat{c}^\pdagger_{i\alpha}
+ U \sum_i \hat{n}_{i \uparrow} \hat{n}_{i \downarrow} \label{HubbardModel}
\end{equation}
on a square lattice of sites, $i$, describing electrons $\hat{c}_{i\alpha}$ with hopping parameters $t_{ij}=t_{ji}\in\mathbb{R}$, chemical potential $\mu$, and on-site repulsion $U$ (summation over Greek indices appearing twice is implied, and $\hat{n}_{i \alpha} \equiv \hat{c}^\dagger_{i\alpha} \hat{c}^\pdagger_{i\alpha}$).
Let us begin by writing the exact path integral of $\hat{H}_U$
in the ``spin-fermion'' form with action $S=S_c+ S_{\text{int}} + S_\Phi$. Using $\tau$ to denote imaginary time, we have ($\beta=T^{-1}$ is inverse temperature)
\begin{equation}
S_c = \int_0^\beta  \diff\tau\left[\sum_{i} c^\dagger_{i\alpha}(\partial_\tau -\mu) c^\pdagger_{i\alpha} - \sum_{i,j} t_{ij} c^\dagger_{i\alpha} c^\pdagger_{j\alpha}  \right]. \label{QuadraticElectronicAction}
\end{equation}
The electrons are coupled locally to a bosonic field $\vec{\Phi}=(\Phi^x,\Phi^y,\Phi^z)$, which describes spin-fluctuations, according to ($\vec{\sigma}=(\sigma_x,\sigma_y,\sigma_z)^T$ are Pauli matrices)
\begin{equation}
S_{\text{int}} = \int_0^\beta  \diff\tau \sum_i c^\dagger_{i\alpha} \vec{\sigma}_{\alpha\beta} c^\pdagger_{i\beta} \cdot \vec{\Phi}_i. \label{SpinFermionCoupling}
\end{equation}
While taking the action for $\vec{\Phi}$ to be $S_\Phi = \frac{3}{2 U} \int_0^\beta  \diff\tau \sum_i \vec{\Phi}_i^2$ leads to an exact representation of the Hubbard model (\ref{HubbardModel}), we will use the more general spin-fermion-model form 
\begin{align}
S_\Phi =  \frac{1}{4g_0}\int_0^\beta  \diff\tau\Bigl[\sum_{i} (\partial_\tau \vec{\Phi}_i)^2 + \sum_{i,j}J_{ij} \vec{\Phi}_i \cdot \vec{\Phi}_j + \sum_i V(\vec{\Phi}_i^2) \Bigr],\label{SpinAction}\end{align}
\change{which we imagine arises as an effective low-energy theory of fluctuating antiferromagnetism from the Hubbard model (\ref{HubbardModel}).}
To describe phases with topological order, we rewrite the path integral as a SU(2) gauge theory by transforming to a `rotating reference frame' \cite{Shraiman88,Schulz90,Schrieffer2004}. To this end, we decompose the electronic fields in terms of the spin and charge degrees of freedom:
\begin{equation}
c_i(\tau) = R_i(\tau) \psi_i(\tau), \quad c^\dagger_i(\tau) =  \psi^\dagger_i(\tau)R^\dagger_i(\tau). \label{RewritingOfOperators}
\end{equation}
Here, the unitary $2\times 2$ matrices $R_i(\tau)$ are the bosonic spinon, and the $\psi_i(\tau)$ the 2-component fermionic chargon operators.
This parameterization introduces an additional redundancy leading to an emergent local $\text{SU}(2)$ gauge invariance \cite{SS09},
\begin{equation}
 R_i(\tau) \,\rightarrow \, R_i(\tau)V_i^\dagger(\tau) , \quad \psi_i(\tau) \,\rightarrow \, V_i(\tau)\psi_i(\tau). \label{SU2GaugeInvariance}
\end{equation}
By design, the transformation in \equref{SU2GaugeInvariance} leaves the electron field operators $c_i(\tau)$ invariant and, hence, is distinct from spin rotation. Note that the chargon field, $\psi$, does not carry spin. In contrast, $R$ carries spin 1/2, and global spin rotations act via {\it left\/} multiplication on $R$ (in contrast to the {\it right\/} multiplication in \equref{SU2GaugeInvariance}).

Inserting the transformation (\ref{RewritingOfOperators}) into the electron-boson coupling $S_{\text{int}}$ and introducing the Higgs field $\vec{H}=(H^x,H^y,H^z)$ via $\vec{\sigma}\cdot\vec{H}_i(\tau) = R_i^\dagger(\tau)\vec{\sigma}R_i(\tau)\cdot\vec{\Phi}_i(\tau)$ we obtain \cite{SS09}
\begin{equation}
S_{\text{int}} = \int_0^\beta  \diff\tau \sum_i \psi^\dagger_{i\alpha} \vec{\sigma}_{\alpha\beta} \psi^\pdagger_{i\beta} \cdot \vec{H}_i. \label{Hpsi}
\end{equation}
Note that the Higgs field, $\vec{H}_i$, transforms 
under the adjoint of the gauge SU(2),
while the spinons and chargons transform as gauge SU(2) fundamentals.
Furthermore, a crucial feature is that the Higgs field does not carry any spin since it is invariant under
global SU(2) spin rotation.

The needed metallic (or insulating) phases with topological order
are obtained simply by entering a Higgs phase where $\langle \vec{H}_i \rangle \neq 0$, while maintaining $\langle R_i \rangle = 0$. Any such phase will preserve spin-rotation invariance, and we will focus on such phases throughout this paper. The vanishing of $\langle R_i \rangle$ arises from large fluctuations in the local rotating reference frame,
and so we are considering states with local magnetic order whose {\it orientation\/} undergoes large quantum fluctuations. However, the magnitude of the local magnetic order remains large, and this is captured by the Higgs field with $\langle \vec{H}_i \rangle \neq 0$.

Depending upon the spatial configuration of $\langle \vec{H}_i \rangle$, different ``flavors'' of topological order with residual gauge group $U(1)$ or $\mathbb{Z}_2$ and potentially also broken discrete symmetries are realized  \cite{SS09,DCSS15b,CS17,OurPreprint}. In this paper, we will focus on the simplest case, with no broken symmetries and $U(1)$ topological order, where the Higgs field resembles AF order,
\begin{equation}
\langle \vec{H}_i \rangle =  (0,0,\eta_i  H_0)^T , \quad \eta_i=(-1)^{i_x+i_y},  \label{HiggsFieldConfig}
\end{equation}
as this scenario has the minimal number of independent parameters (only $H_0$) that can be adjusted to fit the numerical data presented below. 
More complicated Higgs-field configurations, with one or more additional parameters, leading to $\mathbb{Z}_2$ topological order (see \appref{A} for an overview) can be treated similarly, but we will not present explicit results because the current momentum space resolution of our DCA computations and DQMC results do not allow to distinguish between the different phases.


\section{Lines of (approximate) zeros}
\label{sec:NearZeros}
Our central results for the electron spectral functions and the near zeros of the Green's function can be understood qualitatively by considering the effective Hamiltonian $\hat{H}_{\psi}$ for deconfined chargons $\psi$ in the Higgs phase; in the metallic case, this phase corresponds to an 
`algebraic charge liquid' \cite{KKSS08}.
To obtain the effective Hamiltonian, we insert the transformation (\ref{RewritingOfOperators}) into the quadratic electronic action $S_c$ in \equref{QuadraticElectronicAction}. We decouple the resulting quartic hopping $t_{ij} c^\dagger_{i\alpha} c^\pdagger_{j\alpha} =  t_{ij} \psi^\dagger_{i\beta} \bigl(R^\dagger_i\bigr)_{\beta\gamma}\left(R_j\right)_{\gamma\delta} \psi^\pdagger_{j\delta}$
into two quadratic terms, $t_{ij} [ \psi^\dagger_{i\alpha} \left(U_{ij}\right)_{\alpha\beta} \psi^\pdagger_{j\beta} +  \left(\chi_{ij}\right)_{\alpha\beta} \bigl(R^\dagger_iR_j\bigr)_{\alpha\beta}]$,
that are self-consistently related by the mutual mean-field parameters $\left(U_{ij}\right)_{\alpha\beta} = \braket{\bigl(R^\dagger_iR^\pdagger_j\bigr)_{\alpha\beta}}$ and $\left(\chi_{ij}\right)_{\alpha\beta} =  \braket{\psi^\dagger_{i\alpha}\psi^\pdagger_{j\beta}}$. Treating the time-derivative in \equref{QuadraticElectronicAction} in a similar way, we obtain a free effective Hamiltonian $\hat{H}_{\psi}$ that governs the chargon dynamics. 
\change{This corresponds to the phase previously labeled \cite{KKSS08} an algebraic charge liquid (ACL). Further refinements of the theory take into account the binding between the chargons and spinons into electron-like bound states, while retaining the topological order \cite{KKSS08,LVVFL12,PunkSS12,PunkPNAS,SBCS16}: this leads to a fractionalized Fermi liquid \cite{FFL,TSMVSS04,APAV04} (FL*) with small Fermi surfaces. We will not analyze the impact of bound state formation quantitatively in this work.}

\begin{figure}[tb]
\begin{center}
\includegraphics[width=\linewidth]{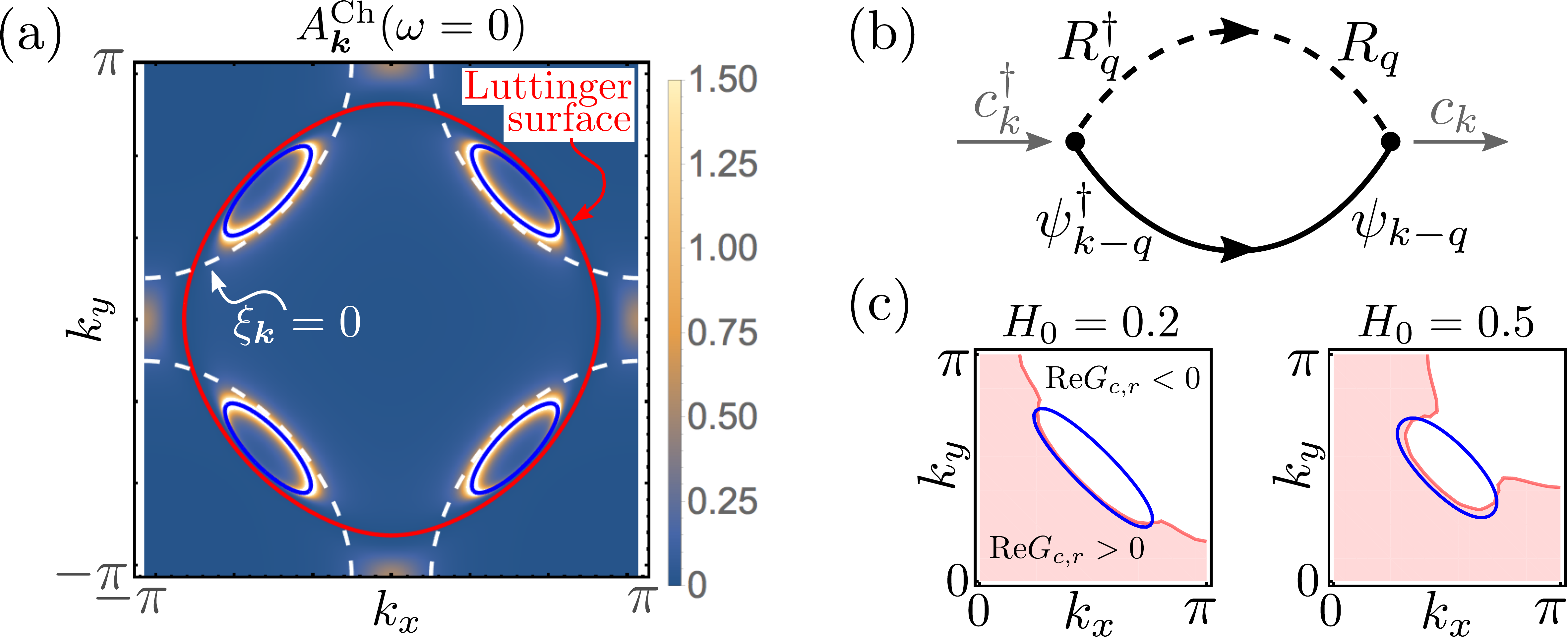}
\caption{(a) The spectral weight $A^{\text{Ch}}_{\vec{k}}(\omega)= -\frac{1}{\pi} \text{Im} G^{\alpha\alpha}_{\psi,\retarded}(\omega,\vec{k})$ associated with the diagonal elements of the retarded chargon Green's function (\ref{GreensFunctionFunction}) is shown at zero energy $\omega=0$ (color plot) together with the Fermi surface of the chargons in the presence (blue line) and absence (white dashed) of the Higgs condensate as well as the Luttinger surface of the chargons (red line). \change{Measuring all energies in units of the nearest neighbor hopping $t$, we have taken $H_0 = 0.5$, $\mu=-0.8$, and a next-to-nearest neighbor hopping of $t'=-0.3$. Throughout this work, all further-neighbor hoppings are taken to be zero for concreteness.} (b) One-loop diagram yielding the electronic Green's function with solid (dashed) lines referring to the chargon (spinon) Green's function. Part (c) shows the reconstructed Fermi surface (blue line) of the chargons and the region in momentum space (light red) where the real part of the retarded electronic Green's function is positive for two different values of $H_0$ \change{as indicated and fixed hole-density of $p=0.1$, $t'=-0.3$, $J^2=0.3$. To describe the zero-temperature limit, we have taken $T$ to be much smaller than all other energy scales.}}
\label{ChargonGreensFunction}
\end{center}
\end{figure}

It can be shown (see \appref{B}) that $U_{ij}$ are trivial in SU(2) space, $U_{ij} = Z_{i-j} \mathbbm{1}$, $Z_{ij}\in\mathbb{R}$, and, hence, only lead to a renormalization of the hopping amplitudes $t_{ij} \rightarrow Z_{i-j} t_{ij}$ inherited from the bare electrons. \change{The chemical potential $\mu$ in \equref{QuadraticElectronicAction} is not renormalized due to the identity $R^\dagger_iR^\pdagger_i=\mathbbm{1}$ and is, thus, identical for electrons and chargons.} Combined with Eq.~(\ref{Hpsi}) we obtain
\begin{align}
\hat{H}_{\psi} = - \sum_{i,j} (Z_{i-j} t_{ij} +\delta_{ij}\mu) \hat{\psi}_{i \alpha}^\dagger \hat{\psi}_{j\alpha}^{\vphantom\dagger}  +\sum_i \hat{\psi}^\dagger_{i\alpha} \vec{\sigma}_{\alpha\beta} \hat{\psi}^\pdagger_{i\beta} \cdot \langle \vec{H}_i \rangle. \label{ChargonHamiltonian}
\end{align}
The key observation is that the Higgs condensate $\langle \vec{H}_i \rangle$ acts just like magnetic order does on the electrons. So the band structure of the $\psi$ fermions is reconstructed into small Fermi surfaces, even though there is no long-range order. For the Higgs configuration in \equref{HiggsFieldConfig},
the momentum-diagonal element of the retarded chargon Green's function is 
\begin{align}\begin{split}
 G_{\psi,\retarded}^{\alpha\beta}(\omega,\vec{k}) &= \frac{\delta_{\alpha\beta}}{\omega + i\eta - \xi_{\vec{k}}-\Sigma^\retarded_{\psi}(\omega,\vec{k})}, \\  \Sigma^\retarded_{\psi}(\omega,\vec{k}) &= \frac{ H_0^2}{\omega + i\eta -\xi_{\vec{k}+\vec{Q}}}, \label{GreensFunctionFunction}
\end{split} \end{align} 
where $\eta \rightarrow 0^+$, $\vec{Q}=(\pi,\pi)$, and $\xi_{\vec{k}}$ is the dispersion inherited from $Z_{i-j} t_{ij}$. The Green's function not only has poles at the reconstructed Fermi surfaces, but also zeros. Within the notation of \equref{GreensFunctionFunction}, the latter are associated with poles of the self-energy $\Sigma$ and occur when $\omega = \xi_{\vec{k}+\vec{Q}}$. In particular, there is a line of zeros $\{\vec{k}|\xi_{\vec{k}+\vec{Q}}=0\}$ at zero energy --- this is the ``Luttinger surface''. The $\vec{k}$ dependence of the chargon Green's function is illustrated in \figref{ChargonGreensFunction}(a) for parameters relevant to the hole-doped cuprates.

Of course, the chargons are not the physical electrons, $c$, (and neither is \equref{GreensFunctionFunction} the full chargon Green's function which also has momentum-off-diagonal components, $\braket{\psi^\pdagger_{\vec{k}}\psi^\dagger_{\vec{k}+\vec{Q}}}$) and so we cannot directly obtain the electron spectral weight from \equref{GreensFunctionFunction}. We have to go back to \equref{RewritingOfOperators}, and compute the physical spectral function after a convolution with that of the fluctuating spinons (see diagram in \figref{ChargonGreensFunction}(b)), 
\begin{equation}
G_{c}(k)= \frac{T}{2}\sum_{\Omega_n} \int_{\text{BZ}}\frac{\diff^2 \vec{q}}{(2\pi)^2}\sum_{\alpha,\beta}   G_R^{\alpha\alpha}(q) G_\psi^{\beta\beta}(k-q), \label{ExpressionForGF}
\end{equation}
which respects all symmetries of the square lattice (see \appref{C}).
Here $G_R$ and $G_\psi$ denote the momentum-diagonal Matsubara Green's function of the spinons and chargons, respectively, $\Omega_n$ are bosonic Matsubara frequencies, and $k=(i\omega_n,\vec{k})$ ($q=(i\omega_n,\vec{q})$) comprise momenta and fermionic (bosonic) Matsubara frequencies. 
\change{While $G_\psi$ follows from \equref{GreensFunctionFunction} upon replacing $\omega + i\eta \rightarrow \omega_n$, it holds
\begin{equation}
G_{R}^{\alpha\alpha}(q) =g\frac{\Omega_n^2 + E^2_{\vec{q}+\vec{Q}}}{\left(\Omega_n^2 + D_{\vec{q}+}^2\right)\left(\Omega_n^2 + D_{\vec{q}-}^2\right)},  \label{SpinonGreensFunctionMainText}
\end{equation}
where we have introduced the two branches, $s=\pm$, of the spinon dispersion
\begin{align}\begin{split}
&D^2_{\vec{q}s} =  \frac{1}{2} \Biggl(E^2_{\vec{q}} + E^2_{\vec{q}+\vec{Q}} + (g\chi_\Omega)^2 \\ & + s\sqrt{(E^2_{\vec{q}}-E^2_{\vec{q}+\vec{Q}})^2 + 2(E^2_{\vec{q}}+E^2_{\vec{q}+\vec{Q}})(g\chi_\Omega)^2 +(g\chi_\Omega)^4 }\Biggr).
\end{split}\end{align}
Here, we have introduced 
\begin{equation}
    E_{\vec{q}} = \sqrt{{\bigl(E^\Phi_{\vec{q}}\bigr)}^2 + {\bigl(E^c_{\vec{q}}\bigr)}^2 + \Delta^2}, \label{EqDisp}
\end{equation}
where $\bigl(E_{\vec{q}}^\Phi\bigr)^2 = -2J^2 (\cos{q_x}+\cos{q_y}-2)$ denotes the part inherited from $S_\Phi$ in \equref{SpinAction}, $\Delta$ the spinon gap, and  $E^c_{\vec{q}}$ is the contribution resulting from the coupling to the chargons.
}

Details of this calculation can be found in \appref{B} and \appref{C} and the resulting retarded electronic Green's function $G_{c,\retarded}(\omega,\vec{k})$ will be compared with DCA and DQMC below. \change{While the full spinons Green's function is used in our calculations, we point out that, for the parameters used in the plots, it is a good approximation to use $G_{R}^{\alpha\alpha}(q) \approx g/(\Omega_n^2 + E_{\vec{q}}^2)$, which identifies $E_{\vec{q}}$ as the spinon dispersion.}

The main qualitative observation in \change{the evaluation of \equref{ExpressionForGF}} is that the electron Green's function can be seen as a ``smeared'' version of the $\psi$ Green's function in the limit where the spinon gap $\Delta$ is the smallest energy scale. In particular, the zeros of the chargon Green's function (or, equivalently the poles of the associated self-energy) in \equref{GreensFunctionFunction}, generally become only approximate zeros (peaks of finite height) of the electron Green's function; Signs of these approximate zeros are also found in DCA and DQMC as discussed below.

\section{Modified Luttinger theorem}
\label{ModifiedLuttinger}
The origin of the modified Luttinger theorem in the deconfined Higgs phase can now be easily described. We compute the electron density by the operator identity
\begin{equation}
\hat{c}_{i\alpha}^\dagger \hat{c}_{i \alpha}^{\vphantom\dagger}
= \hat{\psi}_{i\alpha}^\dagger \hat{\psi}_{i \alpha}^{\vphantom\dagger}\,,
\label{ccpsi}
\end{equation}
and apply the standard Luttinger analysis \cite{Potthoff04,KFG15} to the {\it right\/}
hand side of \equref{ccpsi}. The structure of the complete perturbation theory in the deconfined Higgs phase for the $\psi$ correlator is formally the same as that in a conventional long-range-ordered AF phase  for the $c$ correlator. In the Higgs case, we do have additional interactions from the fluctuations of the gauge field, but the stability of the Higgs phase in the metal implies that these can be treated perturbatively. So we can simply transfer all the Luttinger
theorem arguments in \cite{ACDF} for the case of AF order to the deconfined Higgs phase. These arguments then imply that the $\psi$ Fermi surfaces are small
\ie ~in a Higgs state with fluctuating AF order, the $\psi$ Fermi surfaces 
obey the same Luttinger sum rule as those
of the electron Fermi surfaces in a state with long-range AF order\change{: At zero temperature, the hole density $p$ is given by $p=S_{\text{hole}}-S_{\text{double}}$ where $S_{\text{hole}}$ ($S_{\text{double}}$) denotes the fraction of the Brillouin zone where both bands of the chargon Hamiltonian (\ref{ChargonHamiltonian}) are above (below) the Fermi level. 

We note that the exact operator identity in \equref{ccpsi} is only fulfilled ``on average'', $\braket{\hat{c}_{i\alpha}^\dagger \hat{c}_{i \alpha}^{\vphantom\dagger}}=\braket{\hat{\psi}_{i\alpha}^\dagger \hat{\psi}_{i \alpha}^{\vphantom\dagger}}$, in our mean-field treatment of the gauge theory since the unitarity constraint $R_i^\dagger R_i^\pdagger=\mathbbm{1}$ of the spinons is only treated on average, $\braket{R_i^\dagger R_i^\pdagger}=1$. However, this is sufficient for the modified Luttinger theorem discussed above to hold exactly as it is only a statement about the expectation value of the particle number.

The presence of a finite gap $\Delta \neq 0$ in the spinon spectrum leads to a gap at $\omega=0$ in the electronic spectral function $A_{\vec{k}}(\omega)$ at zero temperature. The vanishing zero-frequency spectral weight of the electrons does not allow us to define an interacting analog of an electronic Fermi surface at zero temperature. Nonetheless, the chargons are described by the effective Hamiltonian (\ref{ChargonHamiltonian}) and, hence, exhibit well-defined Fermi surfaces. This not only allows us to conveniently fix the particle number in the system as explained above, but also shows that the Higgs phase is characterized by Fermi liquid-like charge and thermal transport at low temperatures.

Due to the vanishing imaginary part of the zero-frequency electronic Green's function $G_{c,r}(\omega=0,\vec{k})$ at $T\rightarrow 0$, the aforementioned lines of approximate zeros of $G_{c,r}(\omega=0,\vec{k})$ in the Brillouin become exact. Notwithstanding the presence of electronic Luttinger surfaces, the perturbative Luttinger theorem \cite{LuttingerWard}, relating the particle density $n=1-p$ to the area in $\vec{k}$ space where $\text{Re}\,G_{c,r}(\omega=0,\vec{k})>0$,
\begin{equation}
  n = 2\int_{\text{BZ}}\frac{\diff^2 \vec{k}}{(2\pi)^2}\, \Theta\left(\text{Re}\, G_{c,r}(\omega=0,\vec{k})\right), \label{PerturbLT}
\end{equation}
is violated in the Higgs phase.} As can be seen in \figref{ChargonGreensFunction}(c), the size of the area with $\text{Re}\,G_{c,r}(\omega=0,\vec{k})>0$ changes with $H_0$ at fixed electron density (keeping the area enclosed by the chargon Fermi surface fixed). \change{The violation of \equref{PerturbLT}} is a manifestation of the non-perturbative nature of the Higgs phase. 

\change{This reveals the crucial conceptual differences to the phenomenological Yang-Rice-Zhang (YRZ) theory \cite{YRZ}: While YRZ introduced an ansatz for the electronic Green's function of the pseudogap state that respects the identity (\ref{PerturbLT}), we provide a gauge-theory description of fluctuating antiferromagnetism that shows that the perturbative result (\ref{PerturbLT}) is not required to hold in strongly coupled systems. Due to \equref{PerturbLT}, the YRZ theory requires a fine-tuned position of the Luttinger surface. In our gauge theory, the lines of (approximate) zeros change continuously with system parameters, without obeying the constraint (\ref{PerturbLT}), and are a consequence of the underlying topological order.}

\begin{figure}[tb]
\begin{center}
\includegraphics[width=\linewidth]{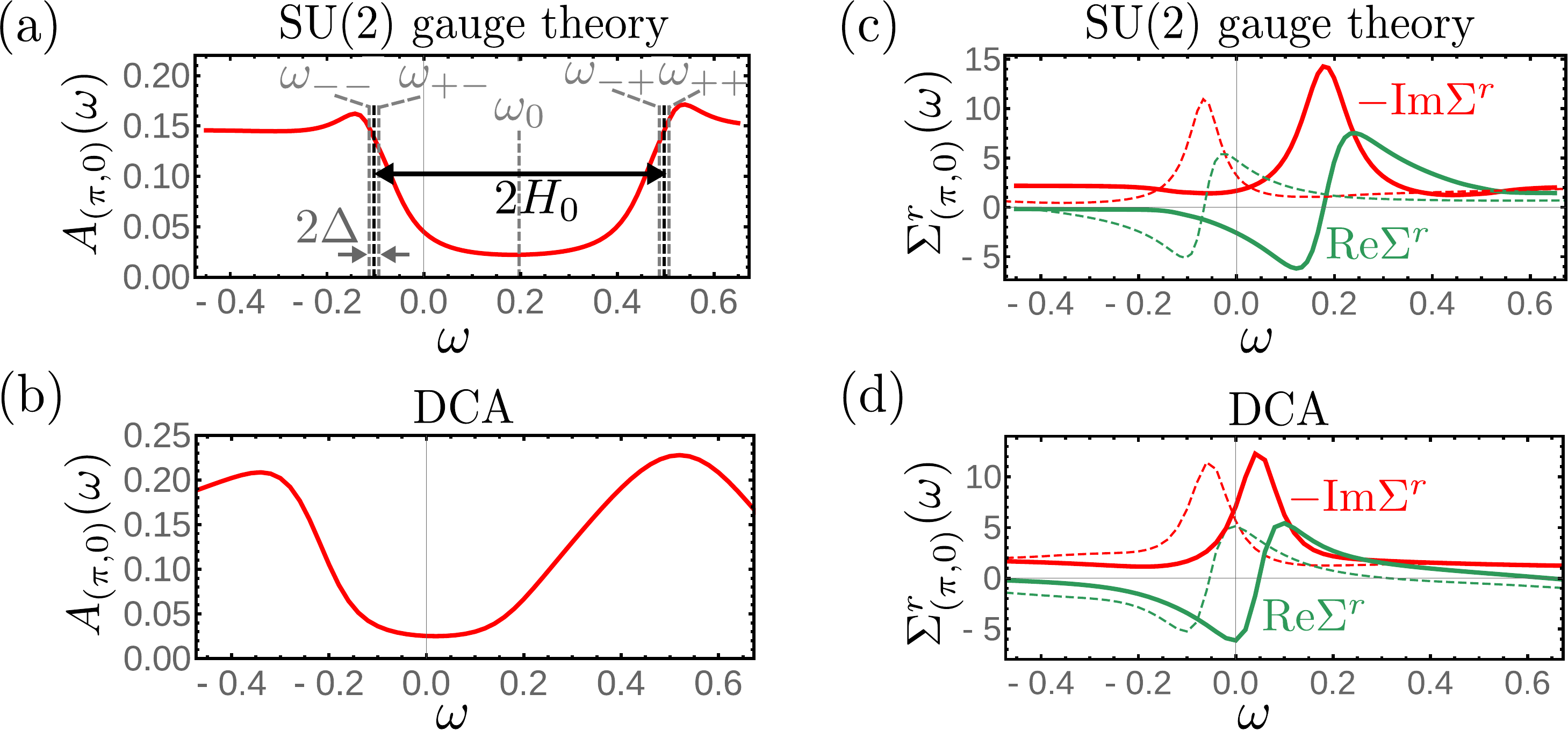}
\caption{Comparison of the electronic spectral weight $A_{(\pi,0)}$ and the retarded electronic self-energy at the anti-nodal point for the SU(2) gauge theory and as obtained in DCA on the Hubbard model. The solid (dashed) lines refer to $t'=-0.15$ ($t'=-0.25$) for DCA and to $t'=-0.15$ ($t'=-0.5$) for the gauge theory model. We used $U=7$, $p=0.05$, and $T=1/30$ for the DCA calculations. In addition, for the gauge theory, we assumed $H_0 =0.3$, $J^2=0.1$, $\Delta=0.01$, and $\eta = 0.04$ in the analytical continuation $i\omega_n \rightarrow \omega + i\eta$ of the gauge theory to cutoff poles in the numerical integration and ``smoothen'' the spectral function.}
\label{MetallicAntinodal}
\end{center}
\end{figure}

\section{Results for the pseudogap metal}
\label{MetallicPhase}
To allow for a direct and systematic comparison of the predictions of the SU(2) gauge theory and the Hubbard model (\ref{HubbardModel}), we have performed DCA and DQMC calculations (see \appref{D} for more details on the numerical methods). 
\change{In the SU(2) gauge theory, the main fitting parameter is the magnitude $H_0$ of the Higgs field, which we choose so as to have a similar size of the anti-nodal pseudogap in DCA and the gauge theory. The spinon gap $\Delta$ is constrained to be of order of or smaller than temperature to allow for zero-frequency spectral weight in the nodal region, as seen in experiment and our numerical calculations. Note that $\Delta$ plays the role of a Lagrange multiplier in the mean-field theory and is, hence, uniquely determined by all other system parameters; in particular, it depends on the spin stiffness $g_0$ in \equref{SpinAction}. However, we take the formally equivalent view of specifying $\Delta$ instead of $g_0$ (which is adjusted accordingly) in the following, as $\Delta$ is physically more insightful in the present context.}
Except for $J$ and $\eta$, which only have minor impact on the qualitative shape of the spectral function (\appref{E}), all other parameters of the SU(2) gauge theory were determined by solving the mean-field equations.
For concreteness, we focus on nearest ($t$) and next-to-nearest neighbor hopping ($t'$). Since we are eventually interested in understanding the pseudogap phase in the hole-doped cuprates, we consider small hole dopings $p>0$ in the regime of large onsite repulsion $U$ (taking $U=7t$ for concreteness). All energies are measured in units of $t$.


\subsection{Anti-nodal point and Lifshitz transition}
The gauge-theory result for the spectral function at the anti-nodal point, $\vec{k}=(\pi,0)$, is shown in \figref{MetallicAntinodal}(a) and displays the strong suppression of the low-energy spectral weight characterizing the pseudogap phase.
To understand this behavior, we first note that the $\vec{q}$-integrand in the expression (\ref{ExpressionForGF}) for the electronic Green's function exhibits poles at the energies $\omega^{\vec{k},\vec{q}}_{ss'} = s E_{\vec{q}} + \ChDisp^{s'}_{\vec{k}-\vec{q}}$ after performing the Matsubara summation and the analytic continuation \change{and using the simplified expression, $G_{R}^{\alpha\alpha}(q) \approx g/(\Omega_n^2 + E_{\vec{q}}^2)$, for the spinon Green's function discussed above. Here $\rho^{s}_{\vec{k}}$ are the two chargon bands ($s=\pm$) of the Hamiltonian $\hat{H}_{\psi}$ in \equref{ChargonHamiltonian} and $E_{\vec{q}}$ the spinon dispersion as given in \equref{EqDisp}.}
In the relevant parameter regime $\Delta < T \ll H_0$, the energies for which the spectral weight is suppressed are determined by the $\vec{q}=0$ components $\omega_{ss'}=\omega^{(\pi,0),\vec{q}=0}_{ss'}$ as can be seen in \figref{MetallicAntinodal}(a) and is discussed in more detail in \appref{C}. Using the explicit form of $E_{\vec{q}}$ and $\ChDisp^{s}_{\vec{k}}$ entering $\omega^{\vec{k},\vec{q}}_{ss'}$, we estimate a gap of size $2H_0$ centered around $\omega_0=\xi_{(\pi,0)}=4Z_{t'}t'-\mu$, where $Z_{t'}$ denotes $Z_{i-j}$ for next-to-nearest neighbors $i$ and $j$. 
The same ``Mott-insulating'' behavior at the anti-nodal point is found in our DCA result shown in \figref{MetallicAntinodal}(b).
By comparison we extract a value of about $H_0 = 0.3$. \change{Note that, while the precise position of the minimum of the spectral function differs in the two approaches, the asymmetry of the peaks in \figref{MetallicAntinodal}(a) and (b) with respect to $\omega=0$ is qualitatively the same. Increasing $t'$ ($t'>-0.15$), the minimum of the anti-nodal spectral function moves towards positive values of $\omega$ in DCA as well.}

For a more detailed comparison of DCA and the gauge theory, we also extract the retarded electronic self-energy $\Sigma^\retarded_{\vec{k}}(\omega)$ from the Green's function, $\Sigma^\retarded_{\vec{k}}(\omega) = -\left(G_{c,\retarded}(\omega,\vec{k})\right)^{-1} + \omega -\epsilon_{\vec{k}}$. 
\change{Here, $\epsilon_{\vec{k}} = -2t(\cos k_x+\cos k_y) -4t' \cos k_x \cos k_y -\mu_0$ is the bare electronic dispersion with $\mu_0$ denoting the bare electronic chemical potential, i.e., in the absence of interactions, $U=0$.}
As can be seen in \figref{MetallicAntinodal}(c) and (d), we also find very good agreement between the gauge theory and DCA for the real and imaginary part of the anti-nodal self-energy $\Sigma^\retarded_{(\pi,0)}(\omega)$: At small negative $t'$ (solid lines), the imaginary part of the self-energy is peaked (and the real part changes sign) at positive energies $\omega=\omega_{\text{peak}}>0$, while $\omega_{\text{peak}}$ changes sign for sufficiently large $-t'$ (see dashed lines). 
The fact that the pseudogap is associated with a peak in the imaginary part of the \change{anti-nodal} self-energy $\Sigma^\retarded_{(\pi,0)}$ (quasi-pole) 
was previously emphasized in several works employing cluster extensions of DMFT \cite{KotliarZeros,maier2002angle,berthod2006,gull2009momentum,sakai2009evolution,gull2010,lin2010}.

Note that the exact value of $t'$ where $\omega_{\text{peak}}$ changes sign differs by a factor of order $1$ in DCA and in the gauge theory. \change{This is expected to be predominantly a consequence of the lack of knowledge of the accurate value of the band-renormalization factors $Z_{i-j}$ due to corrections to $Z_{i-j}$ beyond our mean-field treatment. Furthermore, gauge fluctuations also yield direct corrections to the electronic Green's function, potentially even inducing bound-state formation as already discussed above.}

\begin{figure*}[tb]
\begin{center}
\includegraphics[width=\linewidth]{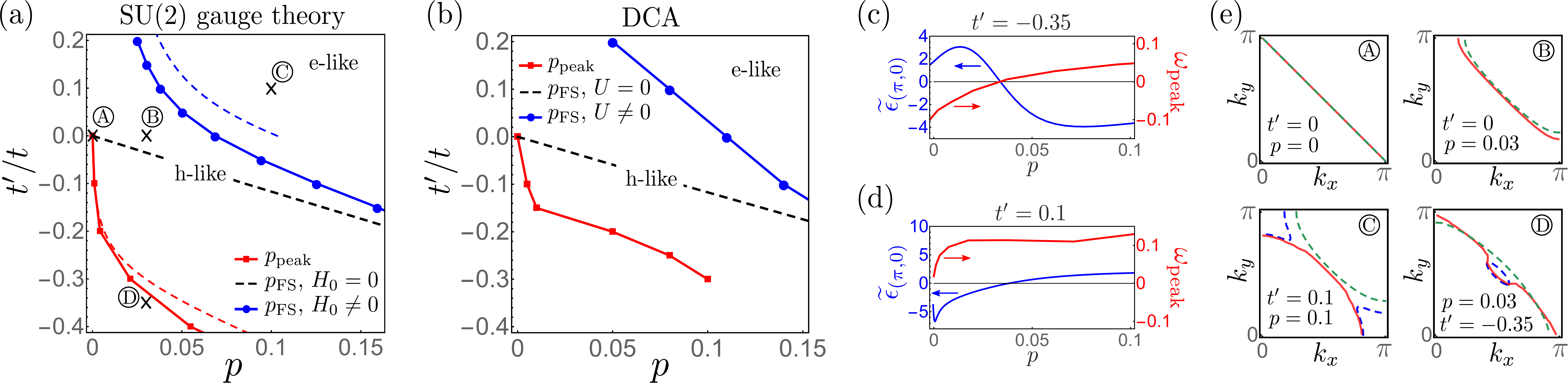}
\caption{The $p$-$t'$ dependence of the ``interacting Lifshitz transition'', defined by the sign change of the renormalized quasiparticle energy $\tilde{\epsilon}_{(\pi,0)}$ at $\omega_{\text{peak}}>0$, is shown as solid blue lines calculated from the SU(2) gauge theory, part (a), and DCA, part (b). The black dashed lines show the location of the same transition for non-interacting electrons. Furthermore, the red lines indicate where the particle-hole asymmetry of the self-energy changes, \ie, where the peak position $\omega_{\text{peak}}$ of the anti-nodal self-energy changes sign. In (a), the dashed blue (red) line corresponds to the parameter configurations where the chemical potential of the chargons touches the top of the lower chargon band (the Lifshitz transition of the chargon Luttinger surface occurs). The doping dependence of $\tilde{\epsilon}_{(\pi,0)}$ and $\omega_{\text{peak}}$ in the gauge theory are shown in (c) and (d) for two different one-dimensional cuts of (a). Finally, (e) shows the position of the zeros of the real part of the electronic zero-frequency Green's function (red solid line), the Luttinger surface (green dashed line) and Fermi surface (blue dashed line) of the chargons for the four distinct points A-D indicated in (a). In this figure, we used $U=7$, $T=1/30$, $H_0 =0.2$, $J^2=0.1$, $\Delta=0.01$, and $\eta = 0.04$.}
\label{MetallicAdditional}
\end{center}
\end{figure*}

More important than exact numerical values, we find the same tendencies of $\omega_{\text{peak}}$ not only as a function of $t'$, but also as a function of hole doping $p$ as summarized in \figref{MetallicAdditional}(a)-(d): The doping value $p_{\text{peak}}$ (red solid line) at which $\omega_{\text{peak}}$ changes sign (at fixed $t'$) increases (from $p_{\text{peak}}=0$ at $t'=0$) with $-t'$ both in the gauge theory, part (a), and in DCA, part (b). Note that, in principle, all parameters of the system depend on doping $p$ and $t'$. However, this dependence is \textit{a priori} unknown. Therefore, we have taken $H_0$ and the spinon Green's function to be constant for all values of $t'$ and $p$ to arrive at \figref{MetallicAdditional}(a) as this requires the fewest number of free tuning parameters and already agrees well with the DCA result. For further discussion we refer to the companion paper \cite{Georges17a}. 

\begin{figure}[bt]
\begin{center}
\includegraphics[width=\linewidth]{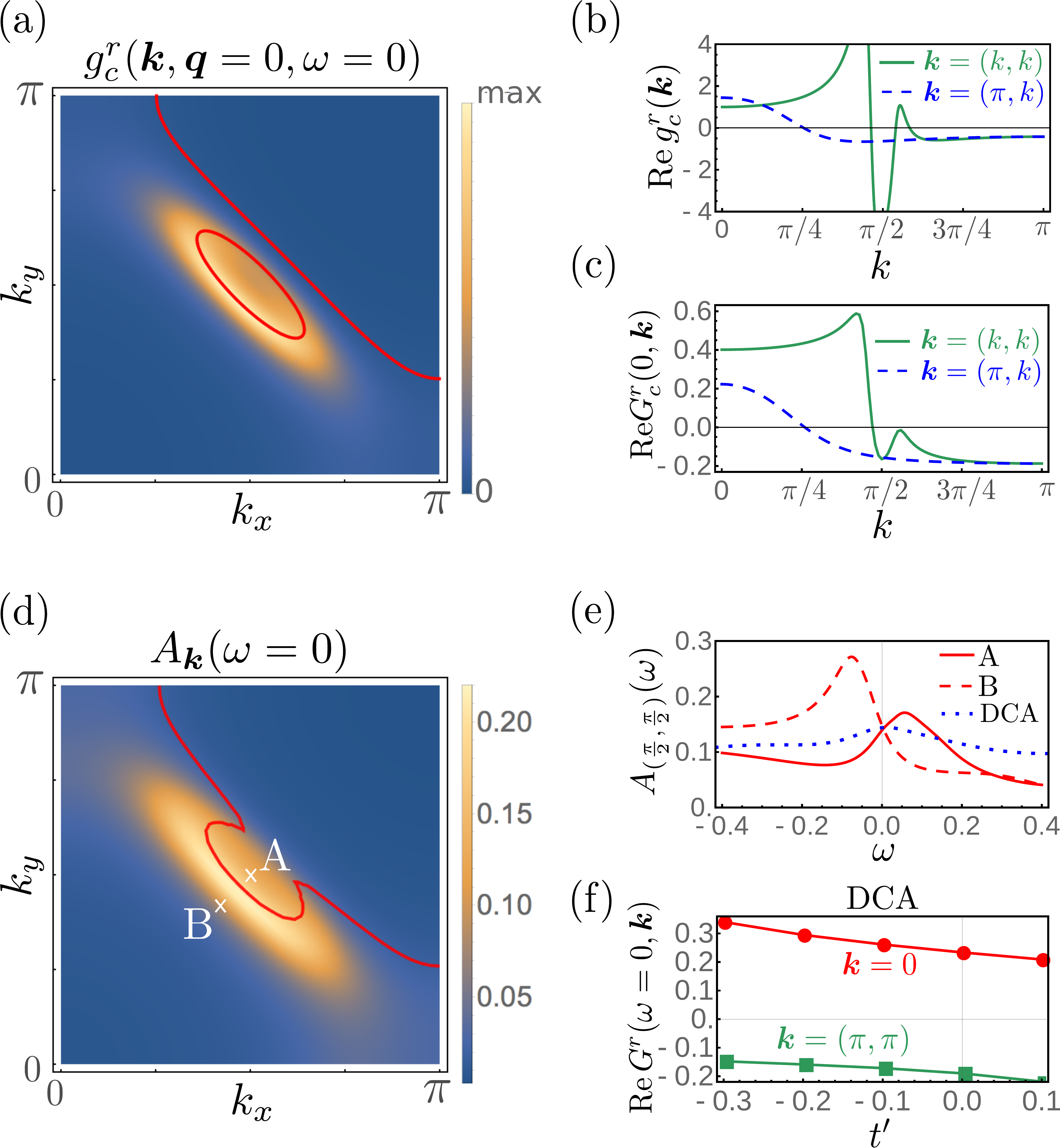}
\caption{In (a), the momentum dependence of the imaginary part (color scale) and the zeros of the real part (red lines) of the $\vec{q}$ integrand, $g_c^\retarded$ are shown. One-dimensional cuts of its  real part, normalized to the value at $\vec{k}=0$, can be found in (b). Part (d) and (c) show the analogous plots for the full electronic Green's function of the gauge theory, \ie, after integration over $\vec{q}$. In (e), the spectral weight at the two momenta A (red solid line) and B (red dashed line) indicated in (d) are shown together with the spectral function of DCA (blue dotted line) averaged over the patch centered around the nodal point $\vec{k}=(\pi/2,\pi/2)$. Part (f) shows the real part of the DCA Green's function at $\vec{k}=(0,0)$ and $\vec{k}=(\pi,\pi)$ as a function of $t'$.
Here we used the same parameters, $U$, $p$, $T$, $H_0$, $J$, $\Delta$, $\eta$, as in \figref{MetallicAntinodal} except for $t'=-0.25$.}
\label{MetallicWholeBZ}
\end{center}
\end{figure}

A simple way to make the doping dependence of $\omega_{\text{peak}}$ plausible proceeds by naively equating the electronic self-energy to that of the chargons in \equref{GreensFunctionFunction} which exhibits a peak at $\omega =\xi_{\vec{k}+\vec{Q}}$. At the anti-node, this frequency vanishes when $\mu(p)=4Z_{t'}t'$. The solution $p=p(t')$ of this equation is shown as a red dashed line in \figref{MetallicAdditional}(a) and closely follows $p_{\text{peak}}(t')$. In this interpretation, the sign change of $\omega_{\text{peak}}$ is associated with the Lifshitz transition of the Luttinger surface of the chargons [cf.~also part B and D in \figref{MetallicAdditional}(e)]. Note that the difference between the Lifshitz transition of the 
chargons' Luttinger surface (given by $\mu(p)=4Z_{t'}t'$) and the non-interacting Lifshitz transition of the electrons [black dashed line in \figref{MetallicAdditional}(a) and (b)], which corresponds to $\mu_0=4t'$, is due to both the renormalization factor $Z_{t'}\neq 1$ and the difference in chemical potentials $\mu_0\neq\mu$ resulting from the reconstruction of the chargon Fermi surface.

We have also studied the renormalized quasiparticle energy,
\begin{equation}
 \tilde{\epsilon}_{\vec{k}} = \epsilon_{\vec{k}} + \text{Re}\,\Sigma^\retarded_{\vec{k}}(\omega=0) = -\text{Re}\left(G_{c,\retarded}(\omega=0,\vec{k})\right)^{-1}, \label{RenormalizeQPEn}
 \end{equation} 
at the anti-nodal point. By analogy to the non-interacting case, the sign change of $\tilde{\epsilon}_{(\pi,0)}$ can be used to define an ``interacting Lifshitz transition'' \cite{Georges17a} between a fictitious hole-like ($\tilde{\epsilon}_{(\pi,0)}<0$) and electronic-like ($\tilde{\epsilon}_{(\pi,0)}>0$) Fermi surface. The location of the ``interacting Lifshitz transition'', $p_{\text{FS}}(t')$, is also indicated (solid blue line) in \figref{MetallicAdditional}(a) and (b). Again, we find good agreement between the gauge theory and DCA. 

A particularly striking aspect of the DCA result (see Ref.~\cite{Georges17a} for more details) is the strong deviation of the interacting Lifshitz transition line (solid blue) to its non-interacting ($U=0$) analogue (black dashed) at low doping. The gauge theory, which shares this feature, admits a qualitative interpretation of this behavior: Starting at $p=0$, $t'=0$, point A in \figref{MetallicAdditional}(a) and (e), and increasing $p$ (at fixed $t'=0$) leads to increasingly negative values of the chemical potential. In the non-interacting limit, the Fermi surface directly becomes \textit{electron}-like. However, in the Higgs phase of the gauge theory, the chargons have a gap and the chemical potential of the chargons stays in the gap for small values of $p$. In this regime, there is no chargon Fermi surface and, hence, the sign of the real part of the chargon Green's function $G_{\psi,\retarded}(\omega=0,\vec{k})$ is solely determined by the position of the Luttinger surface of the chargons. The latter is defined by $\xi_{\vec{k}+\vec{Q}}=0$ and thus becomes more \textit{hole}-like upon reducing the chemical potential (see point B). Viewing the electronic Green's function as a ``smeared'' version of $G_{\psi,\retarded}$ explains why $\text{Re}\,G_{c,\retarded}(\omega=0,\vec{k}=(\pi,0))>0$ and, thus, $\tilde{\epsilon}_{(\pi,0)}<0$ (hole-like) in this regime. Upon further increasing $p$, the chemical potential touches the top of the lower chargon band. For $t'\geq 0$, the resulting chargon Fermi surfaces are located in the vicinity of $(\pi,0)$ changing the sign of the real part of the Green's function at the anti-node (see point C). As expected from this qualitative picture, the line in $p$-$t'$ space where the chemical potential enters the lower chargon band (dashed blue line) roughly follows the interacting Lifshitz transition (solid blue line).


\subsection{Behavior in the full Brillouin zone}
Having established good agreement with DCA results at the anti-nodal point, we can now use the gauge theory to calculate the electronic Green's function in the entire Brillouin zone. 
To gain qualitative understanding of the result, let us first go one step back and investigate the $\vec{q}$-loop integrand $g^\retarded_c$ of the Green's function, defined via $G_{c,\retarded}(\omega,\vec{k}) = \int_{\text{BZ}}\frac{\diff^2 \vec{q}}{(2\pi)^2} g^\retarded_c(\vec{k},\vec{q},\omega)$, that is obtained from \equref{ExpressionForGF} after performing the Matsubara sum and the analytic continuation. For small $\Delta$, its $\vec{q}=0$ component is expected to yield the main contribution to  $G_{c,\retarded}$ and is plotted at $\omega=0$ as a function of $\vec{k}$ in \figref{MetallicWholeBZ}(a). We see that its momentum space structure closely resembles that of the chargons (cf.~\figref{ChargonGreensFunction}). In particular, its real part shows sign changes both at poles (inherited from the Fermi surface of the chargons) and at zeros (stemming from the Luttinger surface of the chargons) as can be more clearly seen in \figref{MetallicWholeBZ}(b).

We expect that lines of zeros of the real part of the Green's function will still be present after $\vec{q}$ integration for sufficiently small spinon gaps, albeit with deformed shape. Indeed, this is what we see in \figref{MetallicWholeBZ}(c) and (d), where the resulting electronic Green's function is shown; Note, however, that the zeros associated with the Luttinger surface and with the Fermi surface of the chargons have merged to a single, ``hybrid'', line of zeros of $\text{Re}\, G_{c,\retarded}(0,\vec{k})$. 

Another consequence of the $\vec{q}$ integration is that it also ``washes out'' the peaks in the spectral function as can be seen by comparing \figref{MetallicWholeBZ}(a) and (d). 
However, for the small value of $\Delta$ assumed in the plot, we recover the metallic, Fermi-arc, behavior in the vicinity of $\vec{k}=(\pi/2,\pi/2)$ in coexistence with the suppression of low-energy spectral weight at the antinode. Depending on whether we consider momenta inside (point A in \figref{MetallicWholeBZ}(d)) or outside (point B) the Fermi arc, the spectral weight is peaked at positive or negative energies (see \figref{MetallicWholeBZ}(e)). This is consistently reflected in the large value of the low-frequency spectral weight at the patch centered around $\vec{k}=(\pi/2,\pi/2)$ of the DCA calculations (blue dotted line).
Note that the much broader distribution of the spectral weight in DCA (with maximum at $\omega=0$) is attributed to the fact that the DCA result is an average over the entire momentum patch around the nodal point $(\pi/2,\pi/2)$. At higher precision, we expect that it is important to account for chargon-spinon interactions in the nodal region, and these could lead to the formation of a FL* state \cite{PunkPNAS}.


We emphasize that the spectral weight is non-zero in the entire Brillouin zone (in the physically relevant regime where temperature is of order of or larger than the spinon gap) and the aforementioned lines of zeros of the real part, which stem in part from the Luttinger surface of the chargons, only correspond to \textit{approximate} zeros of the Green's function. Nonetheless, it provides a natural explanation for the suppression of the spectral weight at the ``backside of the Fermi arc'' as resulting from its proximity to an approximate zero of the Green's function. 

\begin{figure}[tb]
\begin{center}
\includegraphics[width=\linewidth]{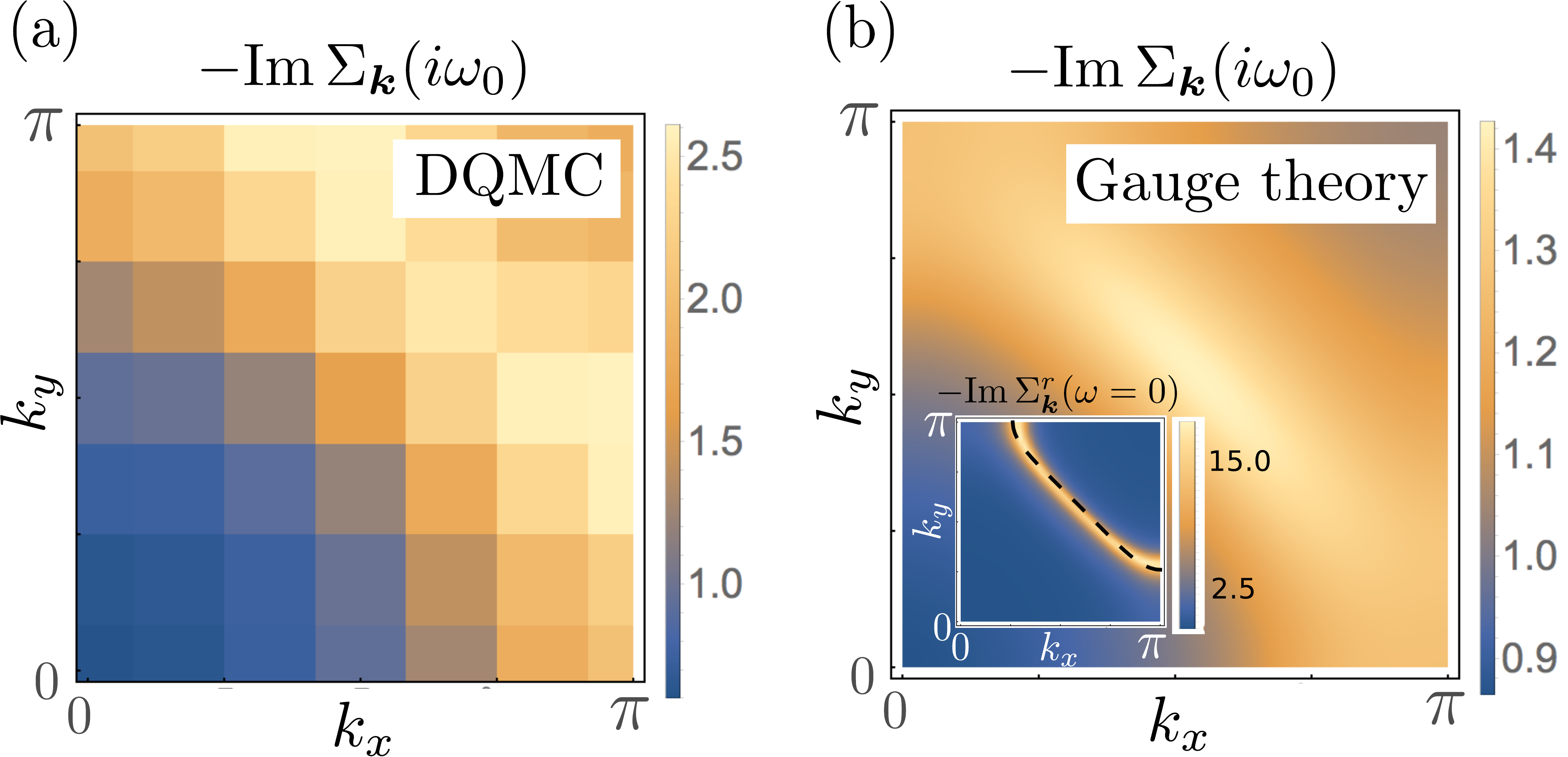}
\caption{The imaginary part of the self-energy at the lowest Matsubara frequency $\omega_0=\pi T$ determined from DQMC on the Hubbard model ($U=7$, $t'=-0.1$, $T=0.25$, $p=0.042$) and from the SU(2) gauge theory is shown in (a) and (b), respectively. The remaining free parameters of the gauge theory have been chosen to be $H_0=0.3$, $J^2=0.1$, $\eta=0.04$, and $\Delta= 0.01 $ as in \figref{MetallicAntinodal}. To avoid too much broadening, we have applied a slightly smaller temperature of $T=0.15$ for the gauge theory. The inset in (b) shows the gauge theory prediction at zero frequency and low temperature (as before $T=1/30$). The black dashed line corresponds to the position of the Luttinger surface of the chargons.}
\label{CompareQMCIm}
\end{center}
\end{figure}

As follows from our discussion, a prediction of the gauge theory, that is robust in the sense that it does not change upon small changes of system parameters, is the sign change of the real part of the low-frequency Green's function from positive at $\vec{k}=0$ to negative at $\vec{k}=(\pi,\pi)$.
We have verified that this sign change is present in our DCA calculations and stable under variation of $t'$ [see \figref{MetallicWholeBZ}(f)]. Note that the sign change cannot be explained in the conventional band description by a hole Fermi surface in the vicinity of the nodal point. \change{The Fermi arc necessitates the presence of a line of sign changes of the real part of $G_{c,r}(\omega=0,\vec{k})$ without a peak in its imaginary part, i.e., an approximate Luttinger surface. In the gauge theory, this additional line of sign changes in the real part is interpreted as} a remnant of the Luttinger surface of the chargons and, hence, indirect evidence for topological order.

We also note that $\vec{k}$-space interpolation schemes of DMFT (see, e.g., \refcite{KotliarZeros,berthod2006,sakai2009evolution,civelli2005dynamical,Kyung_2006,Haule_2007,Kancharla_2008,Liebsch_2009}) 
yield very similar $\vec{k}$-dependence of both the real and the imaginary part of the Green's function as that shown in \figref{MetallicWholeBZ}.
However, instead of interpolating our DCA data, we have also performed DQMC calculations to obtain unbiased and complementary information about the $\vec{k}$ structure of the electronic self-energy. The main limitation of this approach is the sign problem which provides a lower limit for accessible temperatures. Our calculations are performed at $T=0.25 t$ which is of the order of the pseudogap transition temperature estimated from our DCA data \cite{Georges17a}. 
Due to the good qualitative agreement, that we will find below, between the Higgs phase of the SU(2) gauge theory and our DQMC results, we believe that the signatures in the self-energy characteristic of the pseudogap phase are already visible at these temperatures.

The low-energy scattering rate, $\text{Im}\,\Sigma_{\vec{k}}$, (see \appref{F} for the discussion of the real part) obtained from DQMC (evaluated at the lowest Matsubara frequency) is shown in \figref{CompareQMCIm}(a) and has a characteristic peak along an approximately straight line in momentum space. 
We recover this feature in the SU(2) gauge theory as shown in \figref{CompareQMCIm}(b), where it can be interpreted as a consequence of the divergence of the chargon self-energy (\ref{GreensFunctionFunction}) along the chargon Luttinger surface. We have chosen not to perform the analytical continuation of the DQMC data as the mean-field gauge theory can be reliably evaluated both on the real and on the imaginary axis.
\change{Note that we have chosen a slightly reduced temperature for the gauge theory in \figref{CompareQMCIm} as $\text{Im}\,\Sigma_{\vec{k}}$ becomes too ``smeared out'' when taking the value of $T$ used in DQMC. This discrepancy might again be due to the lack of knowledge of the accurate values of $Z_{i-j}$ leading to an effective enhancement of $T$ relative to the chargon energy scales.}

The connection between the peak in the scattering rate and the chargon Luttinger surface can be more clearly seen when calculating $\text{Im}\,\Sigma^\retarded_{\vec{k}}(\omega=0)$ in the gauge theory at low temperature (see inset).
As discussed in \appref{E}, the position of the peak in the scattering rate is found to move closer to $\vec{k}=(\pi,\pi)$ in DQMC upon increasing the value of $U$ \change{around moderate $U \sim 6t$}, and the same trend is found in the gauge theory upon increasing $H_0$.

\section{Summary}
\change{
We have presented a gauge theory of a metal with orientational fluctuations of AF order with a well established local magnitude. In agrement with DQMC results, this theory yields a peak in the electron self energy in the vicinity of the expected location of the Luttinger surface of the charged fermions, while preserving translational symmetry;
the $U$ dependence of the peak location also agrees with DQMC.
In combination with the good agreement between DCA and the gauge theory, this provides evidence for topological order in the hole-doped Hubbard model on the square lattice over the temperature and doping ranges studied. We have compared the numerics with the simplest U(1) Higgsing of a SU(2) gauge theory, but other flavors of topological order remain possible, and more precise studies will be required to refine
the theory.}

\section*{ACKNOWLEDGEMENTS}
{This work was supported by the US National Science Foundation under Grant DMR-1664842, the Simons Foundation Many-Electron Collaboration, the European Research Council (project ERC-319286-`QMAC'), the Swiss National
       Supercomputing Centre (CSCS, project s575),  and MURI grant W911NF-14-1-0003 from ARO. The Flatiron Institute is supported by the Simons Foundation. 
       Research at Perimeter Institute is supported by the Government of Canada through Industry Canada and by the Province of Ontario through the Ministry of Research and Innovation. 
       MS acknowledges support from the German National Academy of Sciences Leopoldina through grant LPDS 2016-12.
       SS acknowledges support from Cenovus Energy at Perimeter Institute, and from the Hanna Visiting Professor program at Stanford University. AG and SS are grateful for the stimulating atmosphere of the May 2017 workshop on cuprates organized by the Institut Quantique in Jouvence, Canada.}


\widetext
\appendix

\section{Topological order in gapless metallic systems}
\label{AppA}
Topological order is usually considered in 
systems with an energy gap to all excitations, where it is defined by a degeneracy of the ground state on a torus \cite{Wen16}, or more precisely, an energy splitting between the lowest energy states of order
$\exp(-\alpha L)$, for some positive constant $\alpha$ as the size of the torus $L \rightarrow \infty$.
Such systems also have emergent gauge fields and `superselection' sectors \cite{Kitaev06} in their excitation spectrum (the superselection sectors are bulk excited states in the infinite system 
which cannot be reached from the ground state by the action of any local operator).

In this work, we are interested in gapless systems, and for such systems, the splitting between lowest
energy states on a torus can be as large as a power of $1/L$ when there are excitations which carry charges
of the emergent gauge field. But
emergent gauge fields and superselection sectors are more robust criteria (and definitions) for the presence of topological order \cite{SSreview18}. In the main text, we analyze
an example of a metallic state with U(1) topological order \cite{Hermele04,KKSS08,EffTheory}, which has an emergent overdamped photon in the excitation spectrum and a superselection sector of `spinon' states. The presence of the photon is directly linked to the suppression of topological  `hedgehog/monopole' defects in the fluctuating antiferromagnetic order \cite{NRSS90}, and so it is appropriate to identify such states as possessing {\it topological} order. 
In the following, we also discuss how metallic states with $\mathbb{Z}_2$ topological order \cite{NRSS91,Wen91}, which have  superselection sectors of spinon and `vison' states, are realized within the SU(2) gauge theory.

\subsection{Different Higgs condensates}
Different flavors of topological order are obtained depending on the texture of the Higgs-field condensate $\langle \vec{H}_i \rangle$. The cases considered previously are \cite{SS09,DCSS15b,CS17,OurPreprint} (the labels are chosen to correspond to those in Ref.~\onlinecite{OurPreprint}):
\begin{enumerate}
\item[(D)]  U(1) topological order with a gapless photon: \begin{equation}\langle \vec{H}_i \rangle = H_0 \, \vec{a} \, \cos (\vec{Q} \cdot
\vec{r}_i), \text{ with } \vec{Q} = (\pi, \pi). \label{HiggsFieldUsed}\end{equation}
\item[(A)] $\mathbb{Z}_2$ topological order with no broken symmetry: \begin{equation}\langle \vec{H}_i \rangle = H_0 \, \vec{a} \, \cos (\vec{Q} \cdot
\vec{r}_i) + H_1 \, \vec{b}, \text{ with } \vec{Q} = (\pi, \pi).\end{equation}
\item[(B)] $\mathbb{Z}_2$ topological and Ising nematic order: 
\begin{align} \langle \vec{H}_i \rangle = &H_0 \left(  \vec{a} \, \cos (\vec{Q} \cdot
\vec{r}_i) +  \vec{b} \,\sin (\vec{Q} \cdot
\vec{r}_i)\right) ,  \text{ with } \vec{Q} \text{ incommensurate.}  \end{align}
\item[(C)] $\mathbb{Z}_2$ topological and current loop order: 
\begin{align}\langle \vec{H}_i \rangle = H_0 &\left(  \vec{a} \, \cos (\vec{Q} \cdot
\vec{r}_i) +  \vec{b} \,\sin (\vec{Q} \cdot \vec{r}_i)\right) + H_1 \, \vec{a}\times\vec{b},  \text{ with } \vec{Q} \text{ incommensurate.}  \end{align} 
\end{enumerate}
Here $\vec{a}$ and $\vec{b}$ are two arbitrary orthonormal vectors satisfying
\begin{equation}
\vec{a}^2 = \vec{b}^2 = 1, \quad \vec{a} \cdot \vec{b} = 1. \label{so3}
\end{equation}
The constants $H_0$ and $H_1$ determine the magnitude of the Higgs condensate (and hence the magnitude of the anti-nodal gap). All 4 phases above preserve translational and spin-rotation symmetry for physical, gauge-invariant observables. 
Phase D, which we have focused on in the main text, and phase A preserve the complete square lattice space group and time-reversal symmetry. Phase B breaks a lattice rotation symmetry, while phase C breaks inversion and time-reversal symmetry, but not their product.

\begin{figure}[tb]
\begin{center}
\includegraphics[width=0.6\linewidth]{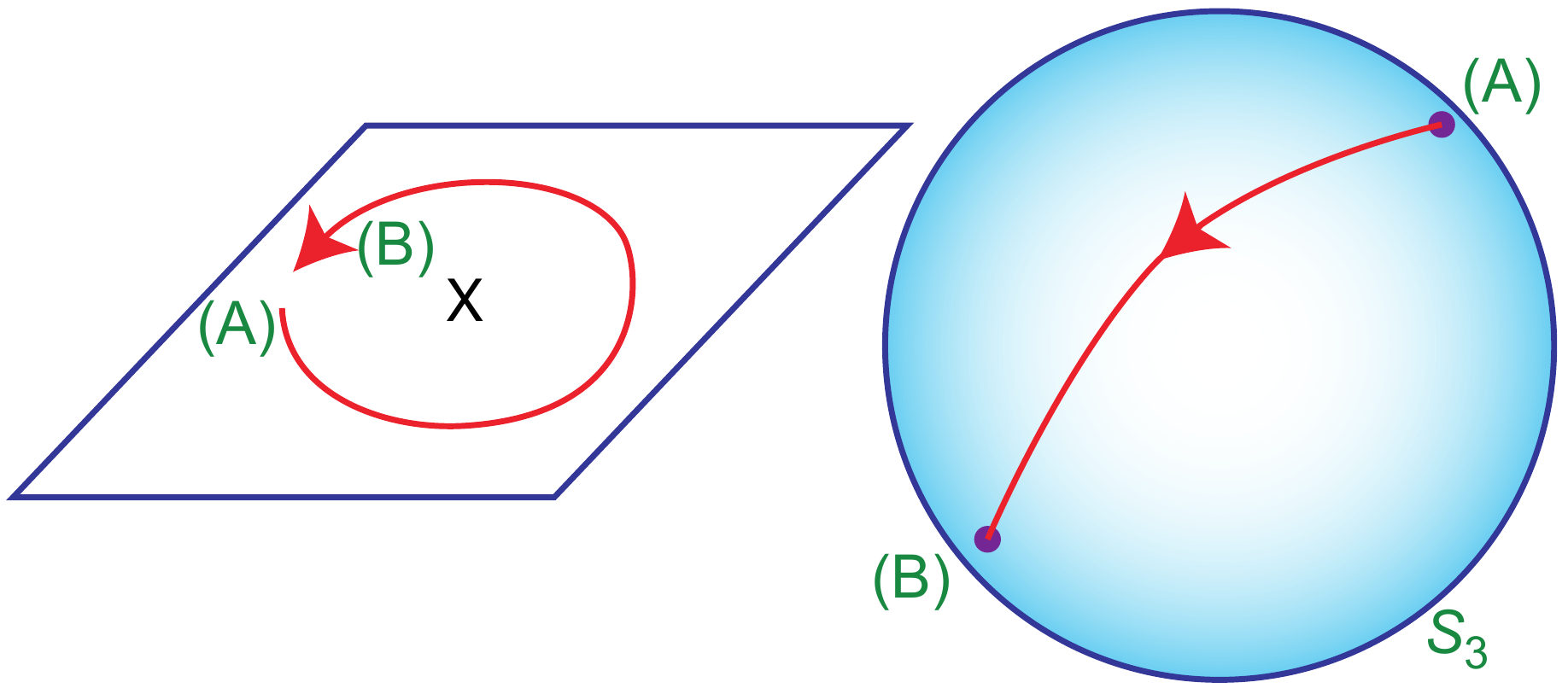}
\caption{The trajectory of $w_\alpha$ on $S_3$ around a vison defect centered at X.
All such anti-podal configurations of $w_\alpha$ are averaged over.}
\label{fig:vison}
\end{center}
\end{figure}

\subsection{Vison states}
We saw above that all three states with $\mathbb{Z}_2$ topological order were characterized by two orthonomal vectors, $\vec{a}$ and $\vec{b}$, obeying Eq.~(\ref{so3}).
The space of such orthonormal vectors is isomorphic to SO(3). The homotopy group $\pi_1 (SO(3)) = \mathbb{Z}_2$ implies that there are stable point-like $\mathbb{Z}_2$ defects: these constitute the vison superselection sector. Because these are defects in the Higgs phase of a gauge theory, 
the energy of each defect is finite (and does not diverge logarithmically with system size, like vortices in an XY model): the SU(2) gauge field screens the gradients of the Higgs field far from the core of the vortex, just as in the case of the Abrikosov vortex in the Ginzburg-Landau theory of a superconductor. So such vison defects are gapped, and they are suppressed in the ground state, thus creating topological order.

More explicitly \cite{CSS94}, 
let us write $\vec{a}$ and $\vec{b}$ in terms of the pair of complex
numbers $w_{1,2}$ via
\begin{equation}
\vec{a} + i\vec{b} = \epsilon_{\alpha\gamma} w_\gamma \vec{\sigma}_{\alpha\beta} w_\beta \,.
\label{s3}
\end{equation}
Then with $|w_1|^2 + |w_2|^2 = 1$, it can be verified that the constraints in Eq.~(\ref{so3})
are automatically satisfied. Note that $w_\alpha$ and $-w_\alpha$ both map to the same
values of $\vec{a}$ and $\vec{b}$. So the mapping in Eq.~(\ref{s3}) is 2-to-1: the complex
number $w_\alpha$ defines the surface of a unit sphere in 4 dimensions, $S_3$, and Eq.~(\ref{s3}) establishes the well-known result SO(3)$ \cong S_3 /\mathbb{Z}_2$. The vison defect is now easy to identify in the $w_\alpha$ parameterization: as one encircles the defect, $w_\alpha$
moves to its anti-podal point; see Fig.~\ref{fig:vison}.

However, one should not think of the $w_\alpha$
as having a definite orientation around the core of the vortex: because of SU(2)
gauge fluctuations, all orientations of $w_\alpha$ are averaged over, while maintaining the anti-podal relation around the core of the vortex.

\section{Electronic Green's function in the SU(2) gauge theory}
\label{AppB}
In this appendix, we describe how the retarded electronic Green's function $G_{c,\retarded}(\omega,\vec{k})$ is calculated within the SU(2) gauge theory.   
To this end, let us first derive an effective spinon-chargon action. Inserting the transformation 
\begin{equation}
c_i(\tau) = R_i(\tau) \psi_i(\tau), \quad c^\dagger_i(\tau) =  \psi^\dagger_i(\tau)R^\dagger_i(\tau) \label{RewritingOfOperatorsSI}
\end{equation}
into the non-interacting part $S_c$ of the electronic action, the hopping terms become (summation over repeated indices is implied)
\begin{align}
 t_{ij} c^\dagger_{i\alpha} c^\pdagger_{j\alpha} =  t_{ij} \psi^\dagger_{i\beta} \bigl(R^\dagger_i\bigr)_{\beta\gamma}\left(R_j\right)_{\gamma\delta} \psi^\pdagger_{j\delta}. \label{RHSofHopp}
\end{align}
We decouple it into two quadratic terms, \ie, replace the right-hand side of \equref{RHSofHopp} by
\begin{equation}
t_{ij} \left( \psi^\dagger_{i\alpha} \left(U_{ij}(\tau)\right)_{\alpha\beta} \psi^\pdagger_{j\beta} +  \left(\chi_{ij}(\tau)\right)_{\alpha\beta} \bigl(R^\dagger_iR_j\bigr)_{\alpha\beta}\right).
\end{equation}
In this approximation, the electronic hopping amplitudes lead to both chargon and spinon hopping terms which are self-consistently related by the mutual mean-field parameters
\begin{subequations}\begin{align}
\left(U_{ij}(\tau)\right)_{\alpha\beta} &= \braket{\bigl(R^\dagger_i(\tau)R_j(\tau)\bigr)_{\alpha\beta}},  \label{UDefinition} \\
\left(\chi_{ij}(\tau)\right)_{\alpha\beta} &=  \braket{\psi^\dagger_{i\alpha}(\tau)\psi^\pdagger_{j\beta}(\tau)}. \label{ChiDefinition}
\end{align}\label{SCMFParam}\end{subequations}
Applying the same procedure to the term involving the time-derivative,
\begin{equation}
c^\dagger_{i\alpha}\partial_\tau c^\pdagger_{i\alpha} \rightarrow \psi^\dagger_{i\alpha} \partial_\tau  \psi^\pdagger_{i\alpha} + \left(\chi_{ii}(\tau)\right)_{\alpha\beta}\bigl(R^\dagger_i \partial_\tau  R^\pdagger_i\bigr)_{\alpha\beta}, 
\end{equation}
we finally obtain the factorized chargon-spinon action $S_\psi + S_R$ where the chargon part is given by 
\begin{align}\begin{split}
S_\psi = \int_0^\beta  \diff\tau\Biggl[\sum_{i} \psi^\dagger_{i}(\partial_\tau - \mu) \psi^\pdagger_{i} - \sum_{i,j} t_{ij} \psi^\dagger_{i} U_{ij} \psi^\pdagger_{j} +  \sum_i \psi^\dagger_{i} \vec{\sigma} \psi^\pdagger_{i} \cdot \vec{H}_i \Biggr] \label{ChargonPart0}
\end{split}\end{align}
and the corresponding spinon action $S_R$ has two contributions, $S_R = S^c_R + S^\Phi_R$. The first one is due to the quadratic part $S_c$ of the electronic action and reads as
\begin{equation}
S^c_R = \int_0^\beta  \diff\tau\,\,\text{tr}\left[\sum_i \chi_{ii}^T R^\dagger_i \partial_\tau R^\pdagger_i - \sum_{i,j} t_{ij} \chi_{ij}^T R^\dagger_iR_j\right], \label{SpinonAction1}
\end{equation}
where $\text{tr}[\dots]$ denotes the trace in SU(2) space, while the second one, $S^\Phi_R$, comprises the contributions stemming from the spin-dynamics encoded in $S_\Phi$ in Eq.~(\eqtx{4}). If we just took $S_\Phi$ of the form $S_\Phi = f(\{\vec{\Phi}_i^2 \})$ (as it results, \textit{e.g.}, from a Hubbard-Stratonovich decoupling of the bare Hubbard interaction), we would get $S^\Phi_R = 0$ since $\vec{\Phi}_i^2 = \vec{H}_i^2$. As discussed in the main text, we here take the more general spin-fermion-model-like form for $S_\Phi$ given in Eq.~(\eqtx{4}).
Applying the transformation (\ref{RewritingOfOperatorsSI}) to this form of $S_\Phi$, we find that the (spatio-temporally) non-local terms lead to a coupling between the Higgs field and the spinons. 
Instead of analyzing the complicated coupled problem, we will assume that the Higgs field resides at one of the saddle-point configurations given in \appref{A}. 
Due to the limited momentum-space resolution of DCA and DQMC, we focus on phase (D) although we have in mind that there are small corrections to this Higgs-field configuration, reducing the residual gauge group from U(1) to $\mathbb{Z}_2$, which are too small to be visible in our numerical calculations.

The resulting $S^\Phi_R$ is very compactly expressed in terms of the N\' eel order parameter $\vec{n}_i =\eta_i \vec{\Phi}_i/H_0$, $\eta_i=(-1)^{i_x+i_y}$, (the normalization by $H_0$ is just for future convenience) upon noting that 
\begin{equation}
\vec{n}_i \cdot  \vec{\sigma} = \vec{a}\cdot R^{\pdagger}_i \vec{\sigma}R_i^\dagger. \label{NRelation}
\end{equation}
One finds
\begin{equation}
S^\Phi_R = \frac{1}{4g}\int_0^\beta  \diff\tau\left(\sum_{i} (\partial_\tau \vec{n}_i)^2 + \sum_{i,j}\eta_i\eta_j J_{ij} \vec{n}_i \cdot \vec{n}_j \right) \label{LatticeNModel}
\end{equation}
where we introduced the rescaled parameter $g=g_0/H_0^2$. 

Having factorized the spin-fermion action into a sum of a spinon part, \equsref{SpinonAction1}{LatticeNModel}, and a chargon part, \equref{ChargonPart0}, which are self-consistently coupled via \equref{SCMFParam}, let us next discuss the chargon and spinon Green's function separately.

\subsection{Chargon Green's function}
\label{AppB1}
To write down the chargon Green's function, let us assume that 
\begin{equation}
U_{ij} = \mathbbm{1} Z_{i-j}, \qquad Z_{i-j}\in\mathbb{R}. \label{SimplifiedFormU}
\end{equation}
Taking $U_{ij}$ to be trivial in SU(2) space and explicitly translation invariant in a gauge in which the Higgs field is given by \equref{HiggsFieldUsed} is an \textit{ad hoc} assumption at this point. However, we will show below that the resulting spinon theory will indeed reproduce this simple form of $U_{ij}$, which proves that there is a self-consistent solution of the coupled spinon-chargon problem with $U_{ij} = \mathbbm{1} Z_{i-j}$. In this case, all symmetries of the square lattice are preserved. 
In principle, there can be further solutions where additional symmetries might be spontaneously broken, but we are not interested in discussing these in the following; Again, the reason is that any symmetry breaking in the pseudogap phase seems to be beyond the current resolution of our numerical calculations.

For notational convenience, let us perform a global gauge transformation leading to $\vec{a} = (0,0,1)^T$ in \equref{HiggsFieldUsed}. With these simplifications, we transform the chargon action (\ref{ChargonPart0}) to frequency-momentum space, $\psi_i(\tau) \rightarrow \psi_{n\vec{k}}$, and write it in quadratic form,
\begin{align}\begin{split}
S_\psi  =   T\sum_{\omega_n}{\sum_{\vec{k}}}'\sum_{\alpha=\pm} \Psi_{n\vec{k}\alpha}^\dagger  \begin{pmatrix} -i\omega_n + \xi_{\vec{k}} & \alpha  H_0 \\ \alpha  H_0 & -i\omega_n + \xi_{\vec{k}+\vec{Q}} \end{pmatrix}  \Psi_{n\vec{k}\alpha}^\pdagger, \label{QuadraticAction}
\end{split}\end{align}
where $\Psi_{n\vec{k}\alpha}=(\psi_{n\vec{k}\alpha},\psi_{n\vec{k}+\vec{Q}\alpha})^T$,  $\vec{Q}=(\pi,\pi)$ and $\xi_{\vec{k}} = -2\sum_{\vec{\eta}} Z_{\vec{\eta}} t_{\vec{\eta}} \cos{(\vec{\eta}\cdot \vec{k})} - \mu$ is the single-particle dispersion (due to translation symmetry, $t_{j+\eta,j} = t_{\vec{\eta}}$). In all plots presented in the main text (and below), we focus on nearest ($t$) and next-to-nearest neighbor hopping ($t'$) with associated renormalization factors denoted by $Z_t$ and $Z_{t'}$, respectively. The prime in the sum over momentum $\vec{k}$ in \equref{QuadraticAction} indicates that we only sum over the reduced Brillouin zone. 

The action and, hence, the Green's function,
\begin{align}\begin{split}
G_{\Psi}^{\alpha\beta}(i\omega_n,\vec{k}) &\equiv -T\braket{\Psi^\pdagger_{n\vec{k}\alpha}\Psi^\dagger_{n\vec{k}\beta}} \\ &= \frac{\delta_{\alpha\beta}}{\left(i\omega_n - \xi_{\vec{k}}\right)\left(i\omega_n - \xi_{\vec{k}+\vec{Q}} \right)- H_0^2} \begin{pmatrix} i\omega_n - \xi_{\vec{k}+\vec{Q}}  & \alpha  H_0 \\ \alpha  H_0 &    i\omega_n - \xi_{\vec{k}} \end{pmatrix}, \end{split} \label{ChargonGreensFuncFin}
\end{align}
are diagonal in SU(2) space which is a consequence of the gauge we haven chosen and will turn out to be very convenient for the following analysis. \equref{ChargonGreensFuncFin} reproduces the diagonal part of the retarded  Green's function in Eq.~(\eqtx{10}) of the main text, with both lines of poles at $\{\vec{k}|\xi_{\vec{k}}=0\}$  (defining the Fermi surface) and lines of zeros at $\{\vec{k}|\xi_{\vec{k}+\vec{Q}}=0\}$ (defining the Luttinger surface) at zero energy.

\subsection{Spinon Green's function}
\label{AppB2}
In order to calculate the Green's function of the spinons, we first rewrite the spinon action in terms of the complex bosonic $\mathbb{C}\mathbb{P}^{1}$ fields $z_{i,\uparrow}$ and $z_{i,\downarrow}$ which are related to $R_i$ according to  
\begin{equation}
R_i = \begin{pmatrix}
z_{i,\uparrow} & -z_{i,\downarrow}^* \\  z_{i,\downarrow} & z_{i,\uparrow}^*
\end{pmatrix}
 \label{SpinonInZ}
\end{equation}
and satisfy the nonlinear constraint $|z_{i,\uparrow}|^2 + |z_{i,\downarrow}|^2 = 1$. Clearly, any SU(2) matrix can be parametrized in this way, however, the notation has been chosen such that \equref{NRelation} (with $\vec{a}=(0,0,1)^T$ in the gauge we have chosen above) is satisfied for $\vec{n}_i = z_i^\dagger \vec{\sigma}z^\pdagger_i$, $z_i = (z_{i,\uparrow},z_{i,\downarrow})^T$, \ie, $z_i^\dagger \vec{\sigma}z^\pdagger_i$ is the N\' eel order parameter --- exactly as in the standard $\mathbb{C}\mathbb{P}^{1}$ description of fluctuating antiferromagnets \cite{CP1}. 

As we will see explicitly below, we have to distinguish two regimes of the spinon-part of the theory: At weak fluctuations, the $\mathbb{C}\mathbb{P}^{1}$ bosons condense, $\braket{z_i} \neq 0$ and $\braket{\vec{n}_i} \neq 0$; Consequently, the system has conventional long-range magnetic order as is realized in the (close to) half-filled Hubbard model. When quantum fluctuations are stronger, the spinons become gapped, $\braket{z_i} = 0$, and $\braket{\vec{n}_i} = 0$; There is no long-range magnetic order, all lattice symmetries are preserved, and the system has U(1) topological order. This is our candidate state for the pseudogap phase in the Hubbard model, which we compare with DCA and DQMC results in the main text.   

Let us begin by analyzing the contribution $S^\Phi_R$ in \equref{LatticeNModel} emanating from the collective antiferromagnetic fluctuations in $S_\Phi$. Taking, \textit{e.g.}, $J_{ij}$ to be finite only on nearest neighbor bonds of the square lattice, $S^\Phi_R$ assumes the familiar relativistic form 
\begin{equation}
S^{\Phi,\text{cont}}_R = \frac{1}{4g} \int\diff^2\vec{r}\diff\tau\left[ (\partial_\tau \vec{n})^2 + v^2  (\partial_{x}\vec{n})^2 + v^2  (\partial_{y}\vec{n})^2  \right] \label{NinCont}
\end{equation}
in the continuum limit.
The corresponding $\mathbb{C}\mathbb{P}^{1}$ model of \equref{NinCont} can be written as \cite{CP1NLsM,EffTheory}
\begin{equation}
S^{\Phi,\text{cont}}_z = \frac{1}{g} \int \diff^2 \vec{r} \diff \tau \, |(\partial_\mu - i a_\mu) z_\alpha|^2,
\end{equation}
where $a_\mu$ is an emergent U(1) gauge field and $\partial_\mu$ comprises spatial and temporal derivatives. In \refcite{OldCalculation}, it was argued that neglecting gauge-field fluctuations will not qualitatively affect the resulting electronic Green's function. Consequently, we will replace $S^\Phi_R$  by the effective lattice $\mathbb{C}\mathbb{P}^{1}$ action
\begin{equation}
S^\Phi_z = \frac{1}{g}T\sum_{\Omega_n}\sum_{\vec{q}} z^\dagger_{n\vec{q}} z^\pdagger_{n\vec{q}} \left[\Omega_n^2 + {\left(E^\Phi_{\vec{q}}\right)}^2 \right],
\end{equation}
where $z_{n\vec{q}}$ are the frequency and momentum transform of $z_i(\tau)$.
Exactly as $J_{ij}$ in Eqs.~(\eqtx{4}) and (\ref{LatticeNModel}), the detailed form of the dispersion $E_{\vec{q}}^\Phi$ is unknown except for the requirement to be periodic in the Brillouin zone and to have a minimum at $\vec{q}=0$. The latter follows from the fact that we are interested in phases in the proximity of antiferromagnetism: When the gap of the $\mathbb{C}\mathbb{P}^{1}$ bosons closes at sufficiently weak quantum \change{fluctuations} (sufficiently small $g$), the $\vec{q}=0$ mode condenses leading to a spatially constant expectation value of $\vec{n}_i$. In the vicinity of $\vec{q}=0$, we expect a relativistic, \textit{i.e.}, linear, energy-momentum relation as already encountered in \equref{NinCont}.   
The arbitrariness in choosing $E_{\vec{q}}^\Phi$ can be resolved by noting that changing the spectrum at high energies is expected to not qualitatively affect the result of the calculation. We, thus, take the simple form $\bigl(E_{\vec{q}}^\Phi\bigr)^2 = -2J^2 (\cos{q_x}+\cos{q_y}-2)$ that meets the aforementioned general requirements.

Let us next consider the second contribution $S^c_R$ to the spinon action in \equref{SpinonAction1} that results from the hopping $S_c$ of electrons on the square lattice. Inserting the parametrization (\ref{SpinonInZ}) and using that $\chi_{ij}(\tau)=\chi_{ij}(0)$ for the chargon action (\ref{QuadraticAction}), $S^c_R$  becomes 
\begin{align}\begin{split}
S^c_z =  \int_0^\beta  \diff\tau \Biggl[&\sum_i \Bigl( (\chi_{ii}^{++} -\chi_{ii}^{--}) z_i^\dagger \partial_\tau z^\pdagger_i + \chi_{ii}^{-+} z_{i\alpha}\epsilon_{\alpha\beta} \partial_\tau z_{i\beta}  - \chi_{ii}^{+-} z^*_{i\alpha}\epsilon_{\alpha\beta} \partial_\tau z^*_{i\beta} \Bigr) \\ & -\sum_{i<j} t_{ij} \Bigl( (\chi_{ij}^{++}+\chi_{ji}^{--})z^\dagger_i z_j  + (\chi_{ij}^{-+}-\chi_{ji}^{-+})z_{i \alpha}\epsilon_{\alpha\beta}z_{j \beta} + \text{c.c.}\Bigr)   \Biggr],\label{SzPre1}\end{split}\end{align}
where  $\epsilon_{\alpha\beta}$ is the Levi-Civita symbol (with $\epsilon_{\uparrow\downarrow}=-\epsilon_{\downarrow\uparrow}=1$) and the shortcut $\chi_{ij}^{\alpha\beta}\equiv (\chi_{ij}(0))_{\alpha\beta}$ is applied. 

Next, we simplify the remaining terms in \equref{SzPre1} further by taking advantage of the structure of the chargon action (\ref{QuadraticAction}) or, equivalently, of the associated Hamiltonian [cf.~Eq.~(\eqtx{9})]
\begin{equation}
  \hat{H}_\psi =   -  \sum_{i,j,\alpha} (Z_{i-j} t_{ij} + \mu \delta_{ij}) \hat{\psi}^\dagger_{i\alpha} \hat{\psi}^\pdagger_{j\alpha}  +  H_0 \sum_{i,\alpha} \alpha \eta_{i} \hat{\psi}^\dagger_{i\alpha} \hat{\psi}^\pdagger_{i\alpha} . \label{ChargonHamiltonian2}
  \end{equation}  
Being diagonal in the SU(2)-index $\alpha$,  $\hat{H}_\psi$ implies $\chi_{ij}^{\alpha\beta} = 0$ if $\alpha\neq \beta$. Consequently, the terms $z_{i\alpha}\epsilon_{\alpha\beta} \partial_\tau z_{i\beta}$ and $z_{i \alpha}\epsilon_{\alpha\beta}z_{j \beta}$ in \equref{SzPre1} vanish in accordance with the symmetry analyis of the $\mathbb{C}\mathbb{P}^{1}$ model in \refcite{OurPreprint}. Furthermore, $\hat{H}_\psi$ has an ``emergent time-reversal symmetry'' (or, in other words, satisfies a reality condition) as it commutes with the antiunitary operator $\hat{\Theta}$ defined by $\hat{\Theta} \hat{\psi}_{i\alpha} \hat{\Theta}^\dagger = \hat{\psi}_{i\alpha}$. We conclude $\text{Im}\, \chi_{ij}^{\alpha\alpha} = \frac{1}{2}\braket{-i \hat{\psi}_{i\alpha}^\dagger\hat{\psi}_{j\alpha}^\pdagger+ i \hat{\psi}_{j\alpha}^\dagger\hat{\psi}_{i\alpha}^\pdagger} = 0$ as the operator $-i \hat{\psi}_{i\alpha}^\dagger\hat{\psi}_{j\alpha}^\pdagger+ i \hat{\psi}_{j\alpha}^\dagger\hat{\psi}_{i\alpha}^\pdagger$ is odd under $\hat{\Theta}$ and, hence, 
\begin{equation}
\chi_{ij}^{\alpha\alpha} = \chi_{ji}^{\alpha\alpha} \in \mathbb{R}. \label{EmTRSConsq}
\end{equation}
In addition, we see that $\hat{H}_\psi$ is invariant under translation by one lattice site and $\alpha \rightarrow -\alpha$ such that 
\begin{equation}
\chi_{i,j}^{\alpha\alpha}=\chi_{i+\vec{e}_\mu,j+\vec{e}_\mu}^{-\alpha-\alpha},\qquad \mu=x,y, \label{TranslationSym}
\end{equation}
and by translation by two lattice sites leading to $\chi_{i,j}^{\alpha\alpha}=\chi_{i+2\vec{e}_\mu,j+2\vec{e}_\mu}^{\alpha\alpha}$. Combining these two observations, we find 
\begin{equation}
\chi_{ii}^{++} -\chi_{ii}^{--} = \chi_{ii}^{++} -\chi_{i+\vec{e}_\mu i+\vec{e}_\mu}^{++} \equiv (-1)^{i_x + i_y}  \chi_\Omega.
\end{equation}
Note that the resulting term in the action,
\begin{equation}
\int_0^\beta  \diff\tau \sum_i  (-1)^{i_x + i_y} \chi_\Omega  \, z_i^\dagger \partial_\tau z^\pdagger_i ,
\end{equation}
is translation invariant as is easily checked by recalling \cite{OurPreprint} that $z_i \rightarrow i\sigma_y z^*_{i+\vec{e}_\mu}$ under translation by one lattice site along the direction $\mu=x,y$.
 
Similarly, \equsref{EmTRSConsq}{TranslationSym} also imply
\begin{equation}
\chi_{ij}^t:=\chi_{ij}^{++}+\chi_{ji}^{--} = \chi_{ij}^{++}+\chi_{i+\vec{e}_\mu j+\vec{e}_\mu}^{++} = \chi_{i-j}^t,
\end{equation}
\ie, the ``spinon-hopping terms'' $t_{ij} \chi^t_{ij}z_i^\dagger z_j$ in \equref{SzPre1} are translation invariant.

With these symmetry-induced simplifications, $S_z^c$ in \equref{SzPre1} assumes the compact form
\begin{align}\begin{split}
S^c_z =  \int_0^\beta  \diff\tau \Biggl[\sum_i (-1)^{i_x + i_y} \chi_\Omega z_i^\dagger \partial_\tau z^\pdagger_i -\sum_{i<j} t_{ij} \chi^t_{i-j} \left(z^\dagger_i z^\pdagger_j  + \text{c.c.}\right)   \Biggr]. \label{SpinonAct2}
\end{split}\end{align}
To make analytical calculations possible, we will treat the nonlinear constraint $z_i^\dagger z_i \equiv |z_{i,\uparrow}|^2 + |z_{i,\downarrow}|^2  = 1$ \change{on quantum and thermal average at each site $i$, \ie, only require}
\begin{equation}
\braket{z_i^\dagger z_i} = 1, \qquad \forall i. \label{MeanFieldConstraint}
\end{equation}
\equref{MeanFieldConstraint} will be accounted for by adding the term $\sum_i \lambda_i z^\dagger_i z^\pdagger_i$ to the action and adjusting the Lagrange multipliers $\lambda_i$ appropriately. \change{Without any further constraint, using a constant Lagrange multiplier $\lambda_i = \lambda$ at each site would imply that \equref{MeanFieldConstraint} holds only on spatial averaging over all sites. However, we note the spinon action $S_z = S_z^c + S_z^\Phi$ is translation invariant (invariant under $z_i \rightarrow i\sigma_y z^*_{i+\vec{e}_\mu}$), and this implies that $\braket{z_i^\dagger z_i}$ is independent of $i$ and \equref{MeanFieldConstraint} holds at each site for a site-independent $\lambda$. }

Performing a transformation to momentum and Matsubara space, we finally obtain the quadratic spinon action
\begin{equation}
S_z = \frac{T}{g}{\sum_{\Omega_n,\vec{q}}}'\sum_{\alpha=\uparrow,\downarrow}Z_{n\vec{q}\alpha}^\dagger  \begin{pmatrix} \Omega_n^2 + E^2_{\vec{q}} & -i\Omega_n g \chi_\Omega \\ -i\Omega_n g \chi_\Omega & \Omega_n^2 + E^2_{\vec{q}+\vec{Q}} \end{pmatrix}  Z_{n\vec{q}\alpha}^\pdagger \label{SpinOnActionFinal}
\end{equation}
with $Z_{n\vec{q}\alpha}=(z_{n\vec{q}\alpha},z_{n\vec{q}+\vec{Q}\alpha})^T$ and the spinon spectrum
\begin{equation}
E^2_{\vec{q}} = {\bigl(E^\Phi_{\vec{q}}\bigr)}^2 + g{\bigl(E^c_{\vec{q}}\bigr)}^2 + \Delta^2, \label{SpinonSpecDef}
\end{equation}
where we have introduced the spinon gap $\Delta$ and the contribution $E^c_{\vec{q}}$ to the spinon dispersion resulting from the coupling to chargons. These two quantities are given by
\begin{align}
\Delta^2 &= g\lambda - 2 g \sum_{\vec{\eta}}t_{\vec{\eta}} \chi_{\vec{\eta}}^t,  \\ {\bigl(E^c_{\vec{q}}\bigr)}^2 &= -2\sum_{\vec{\eta}}t_{\vec{\eta}} \chi_{\vec{\eta}}^t \left(\cos{(\vec{\eta}\cdot \vec{q})}-1\right).
\end{align}
In the following, we will use the more meaningful parameter $\Delta$ instead of $\lambda$ as the quantity that will be adjusted so as to satisfy the constraint (\ref{MeanFieldConstraint}). In the main text, we have absorbed the additional factor $g$ in \equref{SpinonSpecDef} and into $E^c_{\vec{q}}$ to keep the notation short.

In order to calculate the spinon gap $\Delta$, we first need the spinon Green's function which can be read off from \equref{SpinOnActionFinal} to be
\begin{align}\begin{split}
G_{z}^{\alpha\beta}(i\Omega_n,\vec{q}) &\equiv T\braket{Z^\pdagger_{n\vec{q}\alpha}Z^\dagger_{n\vec{q}\beta}} \\
&=\frac{g\,\delta_{\alpha\beta}}{\left(\Omega_n^2 + D_{\vec{q}+}^2\right)\left(\Omega_n^2 + D_{\vec{q}-}^2\right)} \begin{pmatrix} \Omega_n^2 + E^2_{\vec{q}+\vec{Q}}  & i\Omega_n g \chi_\Omega \\ i\Omega_n g \chi_\Omega & \Omega_n^2 + E^2_{\vec{q}} \end{pmatrix},  \label{SpinonGreensFunctionFin}
\end{split}\end{align}
where we have introduced the two branches, $s=\pm$, of the spinon dispersion
\begin{align}\begin{split}
D^2_{\vec{q}s} =  \frac{1}{2} \Biggl(E^2_{\vec{q}} + E^2_{\vec{q}+\vec{Q}} + (g\chi_\Omega)^2 + s\sqrt{(E^2_{\vec{q}}-E^2_{\vec{q}+\vec{Q}})^2 + 2(E^2_{\vec{q}}+E^2_{\vec{q}+\vec{Q}})(g\chi_\Omega)^2 +(g\chi_\Omega)^4 }\Biggr).
\end{split}\end{align}
Upon Fourier transformation, the constraint (\ref{MeanFieldConstraint}) can be rewritten as ($N$ is the number of sites on the square lattice)
\begin{equation}
\frac{T}{N}\sum_{\Omega_n,\vec{q},\alpha} \left[ (G_z^{\alpha\alpha})_{11}(i\Omega_n,\vec{q}) + (G_z^{\alpha\alpha})_{12}(i\Omega_n,\vec{q}) e^{i \vec{Q}\vec{r}_i}   \right] = 1.
\end{equation}
Being odd in Matsubara frequency \change{and convergent ($\propto 1/\Omega_n^3$ as $|\Omega_n| \rightarrow \infty$)}, the second terms vanishes upon summation over $\Omega_n$ as expected from translation symmetry. It shows again that taking a constant $\lambda_i$ was justified. It is straightforward to perform the remaining Matsubara sum which yields ($N\rightarrow \infty$)
\begin{equation}
\int_{\text{BZ}} \frac{\diff^2\vec{q}}{(2\pi)^2} \, f_{\vec{q},T} = \frac{1}{g}, \label{SelfConsistenceCond}
\end{equation}
where the integral goes over the entire Brillouin zone as indicated and
\begin{equation}
f_{\vec{q},T} = \sum_{s=\pm} \frac{\coth\left(\frac{D_s(\vec{q})}{2T}\right)}{D_+^2(\vec{q})-D_-^2(\vec{q})} \frac{D^2_s(\vec{q})-E_{\vec{q}+\vec{Q}}^2}{s D_s(\vec{q})} \label{fqTDefinition}
\end{equation}
has been introduced. 
To arrive at the results shown in the main text, we evaluate the integral over $\vec{q}$ \change{in \equref{SelfConsistenceCond}} numerically. \change{Although $\Delta$ is related to the Lagrange multiplier $\lambda$ and, hence, determined by all other system parameters, we consider (the unknown value of) $g$ rather than $\Delta$ as a free parameter, \text{i.e.}, calculate the value of $g$ for a given value of $\Delta$ by solving \equref{SelfConsistenceCond}.} 

Before proceeding, let us gain intuition for the expected dependence of $\Delta$ on $g$ by considering the limit of small $g \chi_\Omega$ and $T\rightarrow 0$ where \equref{SelfConsistenceCond} reduces to
\begin{equation}
\int_{\text{BZ}} \frac{\diff^2\vec{q}}{(2\pi)^2} \, \frac{1}{E_{\vec{q}}} = \frac{1}{g}. \label{SimplifiedCond}
\end{equation}
From this equation, we can directly see that the integral on the left-hand side decreases with $\Delta$ while being finite at $\Delta = 0$. This shows that $\Delta$ becomes smaller when $g$ is decreased until it vanishes at some critical $g=g_c$ where the transition to a magnetically ordered state occurs. As our central focus is on the pseudogap phase, we only show results for the regime $g>g_c$ where quantum fluctuations suppress magnetic order, no symmetries are broken, and the system has topological order.

Having derived the full spinon action, we can finally come back to the mean-field parameters $U_{ij}$ defined in \equref{UDefinition}. Inserting the transformation (\ref{SpinonInZ}) and taking advantage of the fact that the spinon Green's function (\ref{SpinonGreensFunctionFin}) is diagonal in spin-space, $G_z^{\alpha\beta}\propto \delta_{\alpha\beta}$, we find $U_{ij} = \text{diag}(Z_{ij},Z_{ji})$ with $Z_{ij} = \braket{z_i^\dagger(\tau)z_j^\pdagger(\tau)}$. Upon passing to momentum space, again using that the off-diagonal component of the spinon-Green's function (\ref{SpinonGreensFunctionFin}) is odd in Matsubara frequency, and that $G_z(i\Omega,\vec{q})=G_z(i\Omega,-\vec{q})$, we can finally write 
\begin{equation}
Z_{ij} = g \int_{\text{BZ}} \frac{\diff^2\vec{q}}{(2\pi)^2} \, f_{\vec{q},T} \cos(\vec{q}(\vec{r}_i-\vec{r}_j)) \label{DeterminingZs}
\end{equation}
with $f_{\vec{q},T}$ as defined in \equref{fqTDefinition}. We immediately see that $Z_{ij}=Z_{ji}=Z_{i-j}$ and hence recover the form $U_{ij} =  \mathbbm{1} Z_{i-j}$. This proves that there is a self-consistent mean-field solution with $U_{ij}$ satisfying \equref{SimplifiedFormU}. Due to $f_{\vec{q},T}>0$ and the constraining equation (\ref{SelfConsistenceCond}), we also see that $Z_{i-j} < 1$. 

In order to fully determine the spinon and chargon actions, we have to calculate $\chi_\Omega$ and $\chi_{\vec{\eta}}^t$ from the chargon action (\ref{QuadraticAction}) or, equivalently, from the Hamiltonian in \equref{ChargonHamiltonian2}. To this end, we use $Z_{i-j}=1$. To obtain the chargon action/Hamiltonian, we calculate $Z_{ij}$ from \equref{DeterminingZs} taking $\chi_\Omega=\chi_{\vec{\eta}}^t=1$.

This approximation can be seen as the first iteration in a fully self-consistent calculation of $\chi_{\Omega}$, $\chi^t_{\vec{\eta}}$, and $Z_{i-j}$ where these two sets of parameters are calculated from and reinserted into the chargon and spinon Hamiltonian until convergence is achieved. However, we do not aim to perform a fully self-consistent calculation as the precise value of $\chi_{\Omega}$, $\chi^t_{\vec{\eta}}$, and $Z_{i-j}$ does not change the central results qualitatively and, more importantly, since the parameters of the spinon dispersion $E^\Phi_{\vec{q}}$ inherited from $S_\Phi$ are \textit{a priori} unknown anyway. 


\subsection{Electronic Green's function}
\label{AppB3}
Now we are in position to calculate the electronic Green's function, $G_{c}^{\alpha\beta}(i,j,\tau) = - \braket{c_{i\alpha}(\tau)c^\dagger_{j\beta}(0)}$.
Inserting the transformation (\ref{RewritingOfOperatorsSI}) and using that the effective action derived above is just the sum $S_\psi + S_z$ of a chargon and a spinon contribution without any term coupling the spinons to the chargons directly, we can write
\begin{align}
G_{c}^{\alpha\beta}(i,j,\tau) = -\braket{(R_i(\tau))_{\alpha\alpha'}(R^*_j(0))_{\beta\beta'}}\braket{\psi_{i\alpha'}(\tau)\psi^\dagger_{j\beta'}(0)}. \label{FermionicRealSpace}
\end{align}
While the chargon action $S_\psi$ in \equref{QuadraticAction} seems to break both spin-rotation and translation symmetry, the (matrix) product of chargon and spinon Green's functions determining the electronic Green's function in \equref{FermionicRealSpace} respects both of these symmetries, $G_{c}^{\alpha\beta}(i,j,\tau)=\delta_{\alpha\beta} G_c(i-j,\tau)$, as we show explicitly in \appref{C1} below. 

\begin{figure}[tb]
\begin{center}
\includegraphics[width=0.6\linewidth]{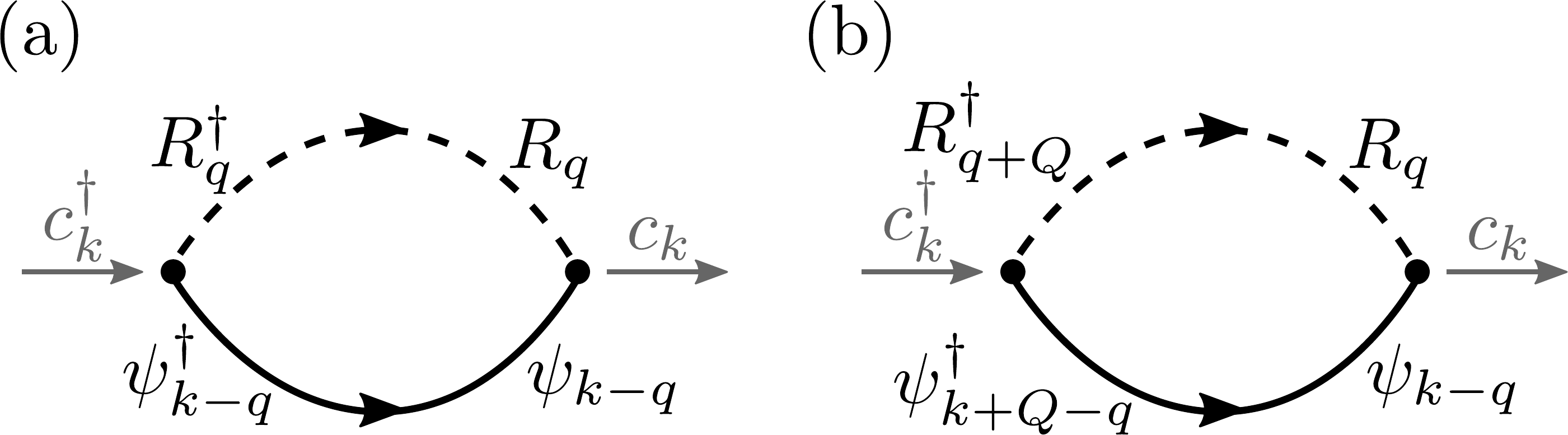}
\caption{Diagrammatic representation of the two contributions to the electronic Green's where the solid (dashed) line refers to the Green's function of the chargons (spinons). We use the compact frequency-momentum notation where $k=(i\omega_n,\vec{k})$, $q=(i\Omega_n,\vec{q})$, and  $Q=(0,\vec{Q})$. As shown in this appendix, the diagram in (b) vanishes.}
\label{Diagrams}
\end{center}
\end{figure}

In momentum-frequency space, \equref{FermionicRealSpace} becomes a convolution with two different contributions related to the fact that both the spinon as well as the chargon action conserve momentum only modulo $\vec{Q}=(\pi,\pi)$. The terms associated with the diagonal and off-diagonal terms of the Green's function are shown diagrammatically in \figref{Diagrams}(a) and (b), respectively.
Using spin-rotation invariance of the spinon action and that $G_z^{\alpha\alpha}(i\Omega_n,\vec{q})$ in \equref{SpinonGreensFunctionFin} is a symmetric matrix, the latter contribution can be written as (comprising frequency-momentum, $q=(\Omega_n,\vec{q})$, $k=(\omega_n,\vec{k})$)
\begin{equation}
\frac{T}{2}\sum_{\Omega_n} \int_{\text{BZ}}\frac{\diff^2 \vec{q}}{(2\pi)^2}\sum_{\alpha,\beta} (G_z^{\alpha\alpha}(q))_{12}(G_\Psi^{\beta\beta}(k-q))_{12}.
\end{equation}
From this expression we directly see that it vanishes identically due to $(G_\Psi^{\alpha\alpha})_{12} = -(G_\Psi^{-\alpha-\alpha})_{12}$ (see \equref{ChargonGreensFuncFin}).  
The electronic Green's function is, hence, entirely determined by the first diagram in \figref{Diagrams}(a), see Eq.~(\eqtx{11}), as stated in the main text.

We next insert the explicit expressions (\ref{ChargonGreensFuncFin}) and (\ref{SpinonGreensFunctionFin}) for the chargon and spin Green's functions into Eq.~(\eqtx{11}) and evaluate the Matsubara sum analytically. 
We can now easily perform the analytic continuation, $i\omega_n \rightarrow \omega + i\eta$, $\eta \rightarrow 0^+$, to obtain the retarded electronic Green's function $G_{c,\retarded}(\omega,\vec{k})$ in form of an integral over the loop momentum $\vec{q}$, 
\begin{equation}
 G_{c,\retarded}(\omega,\vec{k}) = \int_{\text{BZ}}\frac{\diff^2 \vec{q}}{(2\pi)^2} g^\retarded_c(\vec{k},\vec{q},\omega). \label{MomentumIntegral}
\end{equation}
Only the momentum integration in \equref{MomentumIntegral} has to be performed numerically. All results for the electronic Green's function shown in the main text have been obtained without further approximation.

\section{Properties of the electronic Green's function}
\label{AppC}
In this appendix, we discuss further properties of the retarded electronic Green's function obtained from the SU(2) gauge theory as described in \appref{B}.

\subsection{Symmetries}
\label{AppC1}
The translation and spin-rotation invariance of the electronic Green's function is most easily seen in the real space representation (\ref{FermionicRealSpace}): Using the parametrization (\ref{SpinonInZ}) and that both the chargon as well as the spinon Green's function are diagonal in SU(2) and spin space,  $G_\Psi^{\alpha\beta} \propto \delta_{\alpha\beta}$ and $G_z^{\alpha\beta} \propto \delta_{\alpha\beta}$, we can write \equref{FermionicRealSpace} as
\begin{align}\begin{split}
G_{c}^{\alpha\beta}(i,j,\tau) = -\delta_{\alpha\beta}\Bigl(\braket{\psi_{i+}(\tau)\psi_{j+}^\dagger(0)}\braket{z_{\alpha i}(\tau)z^\dagger_{\alpha j}(0)} + \braket{\psi_{i-}(\tau)\psi_{j-}^\dagger(0)}\braket{z_{\overline{\alpha} j}(0)z^\dagger_{\overline{\alpha} i}(\tau)}\Bigr), \label{RealSpaceSymDiscussion}
\end{split}\end{align}
where $\overline{\alpha}=\,\downarrow$ ($\overline{\alpha}=\,\uparrow$) for $\alpha=\,\uparrow$ ($\alpha=\,\downarrow$).
From the spinon action (\ref{SpinOnActionFinal}) further follows that $\braket{z_{\alpha i}(\tau)z^\dagger_{\alpha j}(\tau')}$ does not depend on $\alpha$ and, hence, $G_{c}^{\alpha\beta}(i,j,\tau) \propto \mathbbm{1}_{\alpha\beta}$ leading to a spin-rotation invariant fermionic Green's function. 

As the chargon and spinon actions $S_\psi$ and $S_z$ are invariant under $\psi_{i,\alpha}(\tau) \rightarrow \psi_{i+\vec{e}_\mu ,-\alpha}(\tau)$ and $z_i \rightarrow i\sigma_y z^*_{i+\vec{e}_\mu}$, respectively, we further conclude that
\begin{align}
\braket{\psi_{i,\alpha}(\tau)\psi^\dagger_{j,\alpha}(\tau')} &=\braket{\psi_{i+\vec{e}_\mu,-\alpha}(\tau)\psi^\dagger_{j+\vec{e}_\mu,-\alpha}(\tau')}, \\
\braket{z_{i,\alpha}(\tau)z^\dagger_{j,\alpha}(\tau')} &=\braket{z_{j+\vec{e}_\mu,-\alpha}(\tau')z^\dagger_{i+\vec{e}_\mu,-\alpha}(\tau)}.
\end{align}
Using this in \equref{RealSpaceSymDiscussion}, we find $G_{c}^{\alpha\beta}(i,j,\tau) = G_{c}^{\alpha\beta}(i-j,\tau)$, \ie, the electric Green's function is also translation invariant. 

The remaining square-lattice symmetries (four-fold rotation $C_4$ and mirror reflection $\sigma$) are obviously obeyed by the electronic Green's function as the chargon Hamiltonian (\ref{ChargonHamiltonian2}) is explicitly invariant under these transformations in the gauge we use (more precisely, the gauge transformation $G_i(g)\in\text{SU}(2)$ that accompanies these symmetry transformations $g\in\{C_4,\sigma\}$, with action $\hat{\psi}_{i, \alpha} \rightarrow \sum_{\beta}\left(G_i(g)\right)_{\alpha,\beta} \hat{\psi}_{g(i) , \beta}$, can be chosen to be trivial, $G_i(g)=\mathbbm{1}$, in the present gauge).

Note that the presence of full lattice and spin-rotation symmetry only holds in the limit of sufficiently strong fluctuations (large $g$), where the spinons are gapped, while the condensation of the spinons in the magnetically ordered phase spontaneously breaks spin-rotation and translation symmetry. In this case, also the Green's function becomes nontrivial in spin space and breaks translation symmetry. However, as already mentioned above, this regime is not of interest to us here.

\subsection{Frequency sum rule}
\label{AppC2}
\change{
In this part of the appendix, we prove that the approximations made in \appref{B} for calculating the electron Green's function preserve the electronic frequency sum rule \cite{Abrikosov},
\begin{equation}
\int_{-\infty}^\infty \diff \omega\, A_{\vec{k}}(\omega) = 1. \label{SumRule}
\end{equation}
The validity of the sum rule (\ref{SumRule}) not only constitutes an important consistency check for our approximation scheme, but can also be used to verify that the numerical integration over $\vec{q}$ in Eq.~(11) converges well. Indeed, we have checked that our numerical results for $A_{\vec{k}}(\omega)$ satisfy the sum rule to high precision.

To prove \equref{SumRule}, we will use the field-integral and Green's function description of the main text and \appref{B}; The proof is based on a formal frequency-integration of the expression (11) for the electronic Green's function. Below we will also discuss how this result can be more easily understood from the spectral representation in the operator formalism.
The basic reason for the validity of \equref{SumRule} is that treating the nonlinear constraint $z_i^\dagger z_i^\pdagger$ only on average does not affect the sum rule since the spectral function is related to an expectation value of a quadratic, single-particle, operator.

Before integrating Eq.~(11) over real frequency, we first have to perform the Matsubara sum and the analytical continuation. Using the standard contour integration techniques for Matsubara sums and introducing the shortcuts $G_R^{\vec{q}}(i\Omega_n) \equiv G_R^{\uparrow\uparrow}(i\Omega_n,\vec{q})$, $G_\psi^{\vec{k}}(i\omega_n) \equiv G_\psi^{++}(i\omega_n,\vec{k})$, we have
\begin{align}\begin{split}
G_c(i\omega_n,\vec{k}) = -2 \int_{\text{BZ}}\frac{\diff^2 \vec{q}}{(2\pi)^2}  \Biggl[&  \sum_{\epsilon_B^{\vec{q}}} n_B(\epsilon_B^{\vec{q}}) \text{Res}_{\epsilon_B^{\vec{q}}}[G_R^{\vec{q}}]   G_\psi^{\vec{k}-\vec{q}}(i\omega_n-\epsilon_B^{\vec{q}}) \\ &\quad -\sum_{\epsilon_F^{\vec{k}-\vec{q}}} n_F(-\epsilon_F^{\vec{k}-\vec{q}}) \text{Res}_{\epsilon_F^{\vec{k}-\vec{q}}}[G_\psi^{\vec{k}-\vec{q}}]   G_R^{\vec{q}}(i\omega_n-\epsilon_F^{\vec{k}-\vec{q}})  \Biggr],
\end{split}\label{SumRuleDeriv}\end{align}
where the sums over $\epsilon_B^{\vec{q}}$ and $\epsilon_F^{\vec{k}-\vec{q}}$ involve all poles of $G_R^{\vec{q}}(z)$ and $G_\psi^{\vec{k}-\vec{q}}(z)$, respectively, and $\text{Res}_{z_0}[f]$ denotes the residue of $f(z)$ at $z_0$. In this form, the analytic continuation simply amounts to replacing $G_\psi^{\vec{k}-\vec{q}}(i\omega_n-\epsilon_B^{\vec{q}}) \rightarrow G_{\psi,r}^{\vec{k}-\vec{q}}(\omega-\epsilon_B^{\vec{q}})$ and $G_R^{\vec{q}}(i\omega_n-\epsilon_F^{\vec{k}-\vec{q}}) \rightarrow G_{R,r}^{\vec{q}}(\omega-\epsilon_F^{\vec{k}-\vec{q}})$ with $G_{\psi,r}$ and $G_{R,r}$ denoting the (momentum- and spin-diagonal component of the) retarded chargon and spinon Green's function, respectively. These two Green's functions satisfy
\begin{subequations}\begin{align}
\int_{-\infty}^\infty \diff\omega \, \text{Im} G_{\psi,r}^{\vec{k}}(\omega)  &= -\pi, \\
\int_{-\infty}^\infty \diff\omega \, \text{Im} G_{R,r}^{\vec{q}}(\omega)  &= 0, \label{SpinonSumRule}
\end{align}\label{OtherSumRules}\end{subequations}
where the first identity is just the usual frequency sum rule and the second relation follows from $G_{R}^{\vec{q}}(z)=G_{R}^{\vec{q}}(-z)$ and $(G^{\vec{q}}_R(z))^*=G^{\vec{q}}_R(z^*)$, see Eq.~(12) of the main text. This, in turns, shows that the spectral function $\text{Im}G_{R}^{\vec{q}}(\omega + i \eta)$ is odd in $\omega$ and, hence, the (convergent) integral in \equref{SpinonSumRule} vanishes.  

We further note that both $G_\psi^{\vec{k}}(z)$ and $G_{R}^{\vec{q}}(z)$ are real on the real axis. This implies that also $\text{Res}_{\epsilon_B^{\vec{q}}}[G_R^{\vec{q}}]$ and $\text{Res}_{\epsilon_F^{\vec{k}-\vec{q}}}[G_\psi^{\vec{k}-\vec{q}}]$ are real. In combination with \equref{OtherSumRules}, we find from \equref{SumRuleDeriv} 
\begin{align}\begin{split}
\int_{-\infty}^\infty  \diff\omega \, \text{Im} G^\retarded_c(\omega,\vec{k})  =2\pi \int_{\text{BZ}}\frac{\diff^2 \vec{q}}{(2\pi)^2} \sum_{\epsilon_B^{\vec{q}}} n_B(\epsilon_B^{\vec{q}}) \text{Res}_{\epsilon_B^{\vec{q}}}[G_R^{\vec{q}}].   
\end{split}\end{align}
Rewriting the right-hand side as a Matsubara sum, we finally obtain
\begin{align}\begin{split}
-\frac{1}{\pi}\int_{-\infty}^\infty \diff\omega \, \text{Im} G^\retarded_c(\omega,\vec{k}) = 2T\sum_{\Omega_n}\int_{\text{BZ}}\frac{\diff^2 \vec{q}}{(2\pi)^2} G_R^{\vec{q}}(i\Omega_n) = \braket{z_i^\dagger z_i^\pdagger}=1,
\end{split}\end{align}
which proves the sum rule (\ref{SumRule}).}

As already mentioned above, the sum rule can be more easily understood using the operator formalism.
Applying the standard steps \cite{Abrikosov} for deriving the spectral representation of the Matsubara Green's function ($T_\tau$ denotes time ordering, no summation over $\alpha$),  
\begin{equation}
G_c^{\alpha\alpha}(\tau,\vec{k}) = - \braket{T_\tau \hat{c}^\dagger_{\vec{k}\alpha}(\tau) \hat{c}^\pdagger_{\vec{k}\alpha}(0)}, \qquad \hat{c}_{\vec{k}} = \frac{1}{\sqrt{N}} \sum_{\vec{q}} \hat{R}_{\vec{q}} \hat{\psi}_{\vec{k}-\vec{q}}, \label{OperatorGreensFunction}
\end{equation}
performing the analytic continuation to the retarded Green's function, and integrating the imaginary part of the latter over frequency, one finds  
\begin{equation}
\int_{-\infty}^\infty \diff\omega \,  A_{\vec{k}}(\omega) = \braket{\{\hat{c}^\pdagger_{\vec{k}\alpha},\hat{c}^\dagger_{\vec{k}\alpha} \}}, \label{SumRuleCommu}
\end{equation}
where $\braket{\dots}$ denotes the thermal expectation value.
In systems without fractionalization, 
\begin{equation}
\{\hat{c}^\pdagger_{\vec{k}\alpha},\hat{c}^\dagger_{\vec{k}\alpha} \} = \hat{1}, \label{Anticommutator}
\end{equation}
is usually treated exactly and the sum rule (\ref{SumRule}) follows trivially from \equref{SumRuleCommu}. However, in our case, the electronic operator has been split into spinons and chargons (see  \equref{OperatorGreensFunction}). While the chargon anticommutator is treated exactly, the spinon constraint $z^\dagger_i z^\pdagger_i = 1$ is only treated on average. Consequently, \equref{Anticommutator} does not hold as an operator identity any more, but instead reads as
\begin{equation}
\{\hat{c}^\pdagger_{\vec{k}\alpha},\hat{c}^\dagger_{\vec{k}\alpha} \} = \frac{1}{N} \sum_{\vec{q}} \left( \hat{R}^\pdagger_{\vec{q}} \hat{R}^\dagger_{\vec{q}} \right)_{\alpha\alpha}.
\end{equation}
It is important to note that the frequency sum rule only requires \equref{Anticommutator} to hold ``on average'' which is seen from \equref{SumRuleCommu}. In our approximation scheme described in \appref{B}, this is indeed the case as follows by inserting the $\mathbb{C}\mathbb{P}^{1}$ parameterization (\ref{SpinonInZ}) and transforming back to real space,
\begin{equation}
\braket{\{\hat{c}^\pdagger_{\vec{k}\alpha},\hat{c}^\dagger_{\vec{k}\alpha} \}} = \frac{1}{N} \sum_{i} \braket{\hat{z}_i^\dagger \hat{z}_i^\pdagger} =1.
\end{equation}


\subsection{Modified Luttinger sum rule in the mean-field gauge theory}
A related comment has to be made about the modified Luttinger theorem which is crucially based on the identity 
\begin{equation}
c_{i\alpha}^\dagger c_{i \alpha}^{\vphantom\dagger}
= \psi_{i\alpha}^\dagger \psi_{i \alpha}^{\vphantom\dagger}\,,
\label{app_ccpsi}
\end{equation}
as explained in the main text. While our approach described in \appref{B} does not conserve the identity (\ref{app_ccpsi}) exactly, it is still satisfied on average,
\begin{equation}
\braket{c_{i\alpha}^\dagger c_{i \alpha}^{\vphantom\dagger}}
= \braket{z_i^\dagger z_i^\pdagger}\braket{\psi_{i\alpha}^\dagger \psi_{i \alpha}^{\vphantom\dagger}} = \braket{\psi_{i\alpha}^\dagger \psi_{i \alpha}^{\vphantom\dagger}}.
\end{equation}
Here we have used that the decoupling of the spinon and chargon actions allows to decouple spinon-chargon expectation values into products. For the particle number only the expectation value $\braket{c_{i\alpha}^\dagger c_{i \alpha}^{\vphantom\dagger}}$ matters and, hence, our calculations respect the modified Luttinger sum rule exactly.

\subsection{Simplified expression and anti-nodal spectral weight}
\label{AppC4}
Let us next derive a simplified expression for the electronic spectral weight valid in the limit $g\chi_{\Omega} \rightarrow 0$. In particular, this will help us understand the gauge-theory results shown in Fig.~\eqtx{2} of the main text. We emphasize that this additional approximation has not been made to obtain the results shown in the main text.

Neglecting the off-diagonal term $\propto g\chi_{\Omega}$ in the spinon Green's function (\ref{SpinonGreensFunctionFin}) does not alter the result significantly (it turns out, even at $g\chi_{\Omega} \approx 1$, as we have checked by numerical comparison) as this term only very weakly affects the low-energy part of the spinon spectrum. Performing the same discussion for general $g\chi_{\Omega}$ does not provide any additional physical insights.  

In the limit $g\chi_{\Omega} \rightarrow 0$, the Green's function of the spinons simply reads as $G_z^{\alpha\beta} = g\delta_{\alpha\beta}/(\Omega_n^2+E_{\vec{q}}^2)$ and, hence,
\begin{align}\begin{split}
g_c(\vec{k},\vec{q},i \omega_n) = 2g \, T \sum_{\Omega_n} \frac{1}{\Omega_n^2+E_{\vec{q}}^2} \, \frac{i\omega_n - i\Omega_n - \xi_{\vec{k}-\vec{q}+\vec{Q}}}{(i\omega_n - i\Omega_n - \ChDisp^+_{\vec{k}-\vec{q}})(i\omega_n - i\Omega_n - \ChDisp^-_{\vec{k}-\vec{q}})}, \label{ApproxSpinonGF}
\end{split}\end{align} 
in \equref{MomentumIntegral}. Here we introduced the chargon dispersion
\begin{equation}
\ChDisp^{s}_{\vec{k}} = \frac{\xi_{\vec{k}}+\xi_{\vec{k}+\vec{Q}}}{2} + s \sqrt{\left(\frac{\xi_{\vec{k}}-\xi_{\vec{k}+\vec{Q}}}{2}\right)^2 +  H_0^2}\, .
\end{equation}
As usual, the Matsubara sum can be evaluated using contour deformation and the residue theorem. The resulting expression is a rational function in $i\omega_n$ making the analytical continuation from $g_c(\vec{k},\vec{q},i \omega_n)$ to  $g^\retarded_c(\vec{k},\vec{q},\omega)$ straightforward. The resulting $g^\retarded_c(\vec{k},\vec{q},\omega)$ has poles at four distinct energies
\begin{equation}
\omega^{\vec{k},\vec{q}}_{ss'} = s E_{\vec{q}} + \ChDisp^{s'}_{\vec{k}-\vec{q}}, \qquad s,s'=\pm, \label{PeakEnergies}
\end{equation}
as mentioned in the main text.
While a finite value of $\eta$ is used in the analytic continuation to cutoff the poles for the numerical integration over $\vec{q}$ in \equref{MomentumIntegral} and to introduce a finite life time, we here consider the limit $\eta \rightarrow 0^+$. In this limit, the imaginary part of $g^\retarded_c(\vec{k},\vec{q},\omega)$ and, hence, the electronic spectral weight $A_{\vec{k}}(\omega) = -\frac{1}{\pi} \text{Im} G_{c,\retarded}(\omega,\vec{k})$ can be written in the compact form
\begin{subequations}
\begin{equation}
A_{\vec{k}}(\omega) = \sum_{s,s'=\pm}\int_{\text{BZ}}\frac{\diff^2 \vec{q}}{(2\pi)^2}  Z^{\vec{k},\vec{q}}_{ss'}  \delta(\omega - \omega^{\vec{k},\vec{q}}_{ss'}),  \label{gExpress}
\end{equation}
where the weights are given by
\begin{align}\begin{split}
Z^{\vec{k},\vec{q}}_{ss'} &= g\, n^{\vec{k},\vec{q}}_{ss'}(T) \frac{|\ChDisp^{s'}_{\vec{k}-\vec{q}}-\xi_{\vec{k}-\vec{q}+\vec{Q}}|}{E_{\vec{q}}(\ChDisp^+_{\vec{k}-\vec{q}} - \ChDisp^-_{\vec{k}-\vec{q}})}, \\ n^{\vec{k},\vec{q}}_{ss'}&=n_B(E_{\vec{q}}) + n_F(-s\ChDisp^{s'}_{\vec{k}-\vec{q}}), \label{ThermalFactors}
\end{split}\end{align}\label{AnalyticSpectrWeight}\end{subequations}
with $n_B$ and $n_F$ denoting the Bose and Fermi distribution function, respectively.  

From this expression we can easily see that the spectral weight $A_{\vec{k}}(\omega)$ vanishes at zero frequency $\omega=0$ in the entire Brillouin zone in the limit where temperature $T$ is much smaller than the spinon gap $\Delta$: The delta function in \equref{gExpress} leads to $s E_{\vec{q}} = -\ChDisp^{s'}_{\vec{k}-\vec{q}}$ at zero frequency which allows to simplify the thermal factors, $n^{\vec{k},\vec{q}}_{ss'} \rightarrow n_B( E_{\vec{q}}) + n_F( E_{\vec{q}})$, in \equref{ThermalFactors}. Consequently, $A_{\vec{k}}(\omega=0)$ vanishes exponentially in the limit $T \ll \Delta$. This is expected as any zero-energy electronic excitation must necessarily involve the thermal excitation of a spinon with minimal energetic cost of $\Delta$. For the pseudogap metal state, we thus focus on the parameter regime where the spinon gap is smaller or at most of order of temperature, $\Delta < T$. This allows to have finite spectral weights at zero energy in the nodal part of the Brillouin zone as seen in numerical studies of the Hubbard model and in experiments.

We can now also qualitatively understand the anti-nodal spectral function shown in Fig.~\eqtx{2}(a) of the main text:
It holds $\ChDisp_{\vec{k}}^- < 0$ and $\ChDisp_{\vec{k}}^+ > 0$ at and in the vicinity of the anti-nodal point $\vec{k}=(\pi,0)$. While there is a region in the vicinity of $(\pi/2,\pi/2)$ where $\ChDisp_{\vec{k}}^- > 0$, which also contributes to the spectral weight at the anti-node due to the integration over the loop-momentum $\vec{q}$ in \equref{gExpress}, its contribution requires large $\vec{q}$, of order $(\pi/2,\pi/2)$, where the spinon energy $E_{\vec{q}}$ is large; Therefore, this contribution is suppressed. Furthermore, in the limit $H_0 \gg T$ (where $|\ChDisp_{\vec{k}}^s| \gg T$ for $\vec{k}$ near the anti-node), the thermal factors read as
\begin{equation}
n_{++} \sim n_{--} \sim 1 + n_B(E_{\vec{q}}), \qquad n_{+-} \sim n_{-+} \sim  n_B(E_{\vec{q}}).
\end{equation}
We see that the contribution of $\omega^{\vec{k}=(\pi,0),\vec{q}}_{ss'}$ with $s\neq s'$ to the low-frequency ($|\omega| \ll H_0$) spectral function is suppressed due to the thermal factors ($\omega^{\vec{k}=(\pi,0),\vec{q}}_{s-s}=0$ requires $E_{\vec{q}}$ of order $H_0$). Note further that the magnitude $|\omega^{\vec{k}=(\pi,0),\vec{q}}_{ss}|$ of the frequency of the other two poles (with $s=s'$) is a growing function of $E_{\vec{q}}$. We thus expect a suppression of the anti-nodal spectral weight in the range of frequencies $\omega$ with $\omega_{--},\omega_{+-}<\omega<\omega_{-+},\omega_{++}$ where $\omega_{ss'}=\omega^{(\pi,0),\vec{q}=0}_{ss'}$. This agrees well with Fig.~\eqtx{2}(a) of the main text. We also refer to \figref{AntinodalDep} where these four frequencies are shown (as dots) together with the corresponding anti-nodal spectral function for many different values of the system parameters.

\section{Numerical methods}
\label{AppD}
The multi-site DMFT~\cite{georges1996review} result is obtained 
by doing an eight-site dynamical cluster approximation (DCA) \cite{maier2005review},
which captures the physics of short-range spatial correlations and has the capability 
to distinguish the nodal and anti-nodal physics of the pseudogap \cite{Georges17a}. The effective impurity 
cluster mode of DCA is solved by the Hirsch-Fye quantum Monte Carlo method \cite{hirschfye}, where
an imaginary-time step $\Delta \tau  = 1/21$ is used in the Trotter decomposition. In order to solve
the DCA equations self-consistently, we use typically 50 iterations to get convergence.

The unbiased determinant quantum Monte Carlo (DQMC) \cite{bss1981} simulations are performed on a $12\times 12$ lattice with periodic
boundary conditions at a temperature $T=1/4$. The imaginary time step was set to $\Delta \tau =  1/20$ where the discretization errors are negligible. In the DQMC simulations we use $3\times 10^{7}$ Monte Carlo sweeps to collect the data after $1\times 10^{4}$ warmup sweeps.

\section{Further parameter dependencies}
\label{AppE}
In this appendix, we provide additional plots illustrating the parameter dependence of the anti-nodal spectral function in the gauge theory and compare the interaction dependence with the Higgs-field dependence of the scattering rate in DQMC and the gauge theory, respectively.

\begin{figure}[tb]
\begin{center}
\includegraphics[width=0.90\linewidth]{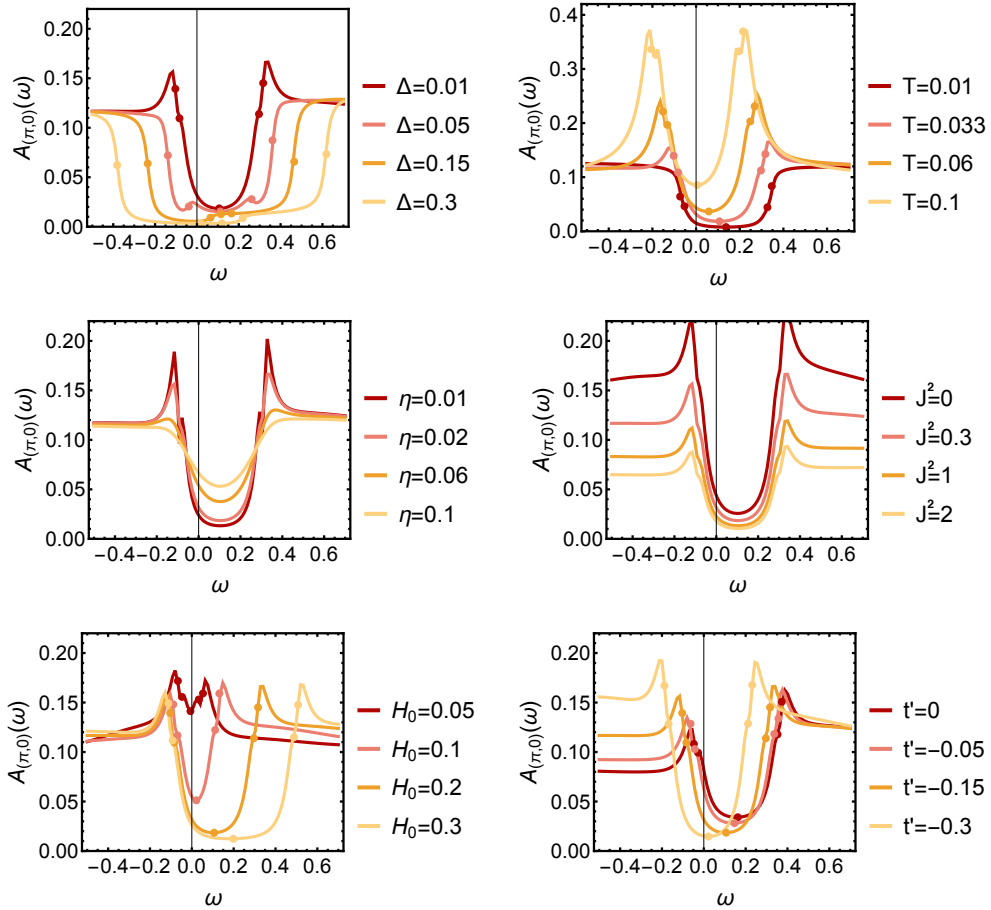}
\caption{Dependence of the anti-nodal spectral function $A_{(\pi,0)}(\omega)$ on various parameters of the system. Unless specified otherwise, we use $t'=-0.15$, $p=0.05$, $T=1/30$, $H_0 =0.2$, $J^2=0.3$, $\Delta=0.01$, and $\eta = 0.02$. The horizontal positions of the dots are given by the five frequencies $\omega_{ss'}=\omega^{(\pi,0),\vec{q}=0}_{ss'}$, $s,s'=\pm$, see \equref{PeakEnergies}, and $\omega_0=\xi_{(\pi,0)}$, while their color and vertical positions indicate which curve the respective point corresponds to.}
\label{AntinodalDep}
\end{center}
\end{figure}

\begin{figure}[tb]
\begin{center}
\includegraphics[width=0.80\linewidth]{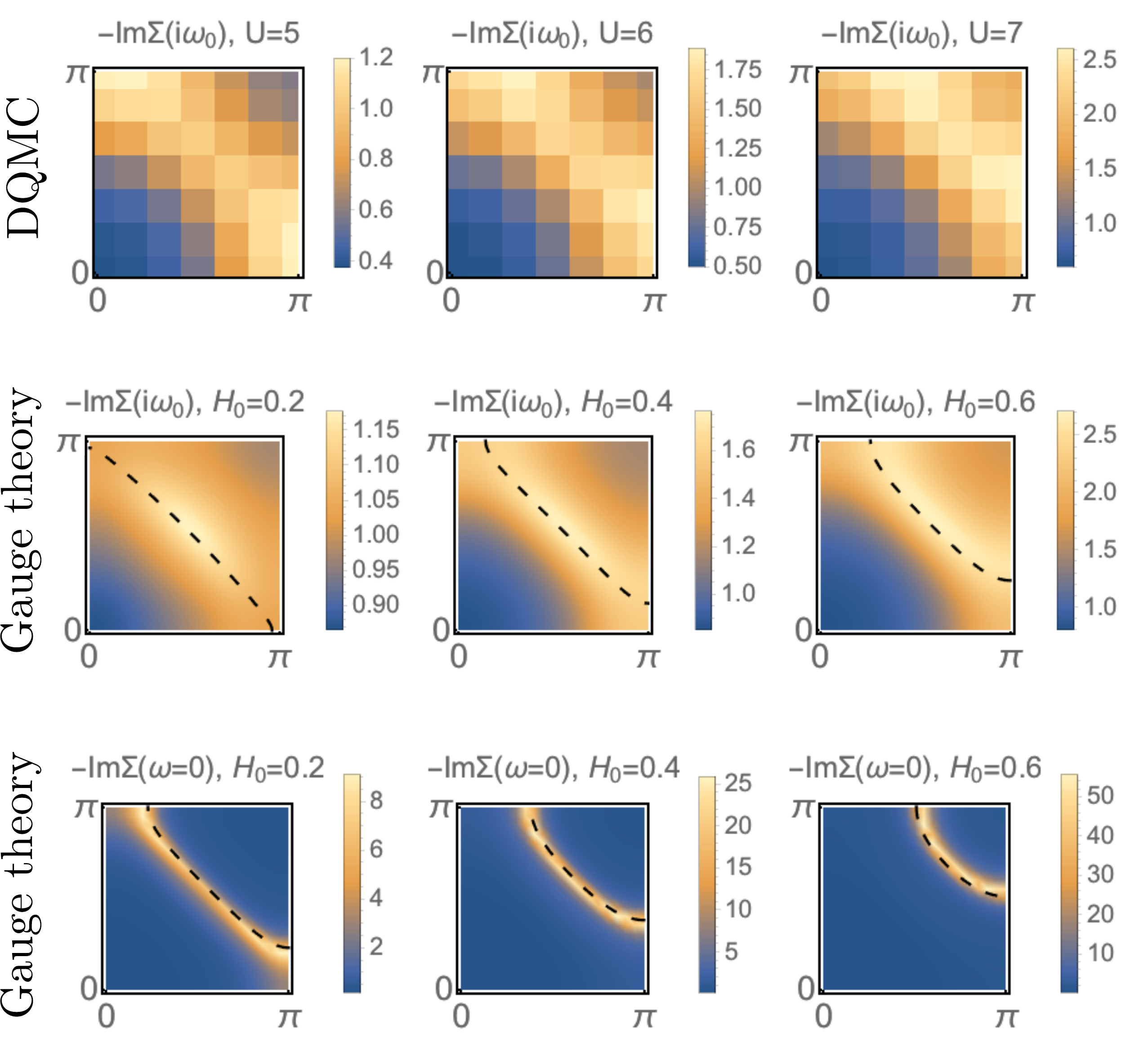}
\caption{Comparison of the interaction ($U$) dependence of the DQMC scattering rate (first line) at $T=0.25$ and the Higgs-field dependence of the gauge-theory self energy evaluated both at high temperatures ($T=0.15$) at the first Matsubara frequency (second line) and at low temperatures ($T=1/30$) at zero frequency (third line). The black dashed lines denote the position of the Luttinger surface of the chargons (at the corresponding temperature). All other parameters are as in Fig.~\eqtx{5} of the main text.}
\label{DifferentUDQMC}
\end{center}
\end{figure} 

\subsection{The anti-nodal spectral weight}
\label{AppE1}
In \figref{AntinodalDep}, the changes in the anti-nodal spectral function $A_{(\pi,0)}(\omega)$ of the gauge theory upon varying any of the $6$ independent parameters, $\Delta$, $T$, $\eta$, $J$, $H_0$, and $t'$, are illustrated (at fixed doping). In accordance with our discussion in \appref{C4}, the four frequencies $\omega_{ss'}$ (indicated as dots) determine the low-energy regime of suppressed spectral weight, \textit{i.e.}, the size of the pseudogap (roughly given by $2H_0$) and its particle-hole asymmetry: The spectral weight is minimal approximately at $\omega=\omega_0:=(\omega_{++}+\omega_{--})/2=\xi_{(\pi,0)}$.   
We further notice (see plots in first line) that $A_{(\pi,0)}(\omega=0) \rightarrow 0$ as $T/\Delta$ approaches zero --- again, as expected from our general discussion in \appref{C4}.

In general, we also see that the height of the peaks in the spectral weight increases with increasing $T$ and decreasing $J$. 
As already discussed in the main text, the main consequence of changing $t'$ is a shift in the minimum of the spectral weight which simply follows from $\omega_0=\xi_{(\pi,0)}=4Z_{t'}t'-\mu$. The main role of $\eta$ is to make the features in the spectral function smooth and, hence, also affects the peak height.

When we compare the gauge theory to the DCA result in Fig.~\eqtx{2}, the main ``fitting'' parameter is the Higgs field strength $H_0$ which is set by (half of) the anti-nodal gap size. The other parameters are either explicitly determined by the parameters used in DCA (like $p$, $t'$, and $T$) or strongly constrained, like $\Delta$ which has to be smaller than $T$ (as discussed in \appref{C4}), or only lead to minor modifications in the overall shape of the spectral function (like $\eta$ or $J$). The values for $\eta$ and $J$ have been chosen in the main text to approximately reproduce the same peak heights in the anti-nodal spectral weight and self-energy as in DCA, see Fig.~\eqtx{2}. In all other comparisons with DCA and DQMC, we used the same values for $\eta$ and $J$. 

We finally note that $g$, $Z_{ij}$, and $\chi_{ij}$ are not additional independent fitting parameters: As explained in \appref{B} in more detail, $g$ is related to $\Delta$ via \equref{SelfConsistenceCond}. We chose to specify $\Delta$ instead of $g$ in our plots due to its physical meaning as the gap of the spinons. 
The matrix elements $\chi_{ij}$ (see \equref{ChiDefinition}) are calculated form the chargon Hamiltonian (\ref{ChargonHamiltonian2}) and the renormalization factors $Z_{ij}$, defined in \equref{UDefinition}, of the chargon Hamiltonian follows from the spinon action according to \equref{DeterminingZs}.

\subsection{Interaction dependence of scattering rate}
\label{AppE2}
\figref{DifferentUDQMC} extends the comparison of the low-energy scattering rate in DQMC and the gauge theory of the main text (cf.~Fig.~\eqtx{5}) to three different values of the Hubbard interaction $U$ and the Higgs field strength $H_0$, respectively. We find that the line of maximal low-energy scattering moves closer to $(\pi,\pi)$ in momentum space both with increasing $U$ and with increasing $H_0$. Within the gauge theory, this shift in momentum space can be seen as a consequence of the shift in position of the Luttinger surface of the chargons.

 \change{Note that for moderate $U \sim 6t$, the paradigm of fluctuating antiferromagnetic order in a metal is assumed to be valid. Therefore, a monotonically increasing relation between the Hubbard $U$ and the resulting Higgs-field strength $H_0$ is consistent with the expectation that a stronger $U$ should increase the magnitude of the local magnetic order. However, at very large $U \gtrsim 10t$ (say), we do expect to move to the Mott limit where the system can be better described by a $t-J$ model with $J \sim t^2/U$. In this limit, increasing $U$ will decrease $J$ which sets the size of the anti-nodal gap at the mean-field level.}

\section{Momentum dependence of the real part of the self-energy}
\label{AppF}
In this appendix, we present the comparison of the real part of the self-energy of the SU(2) gauge theory and of our DQMC data on the Hubbard model.

\begin{figure}[tb]
\begin{center}
\includegraphics[width=0.6\linewidth]{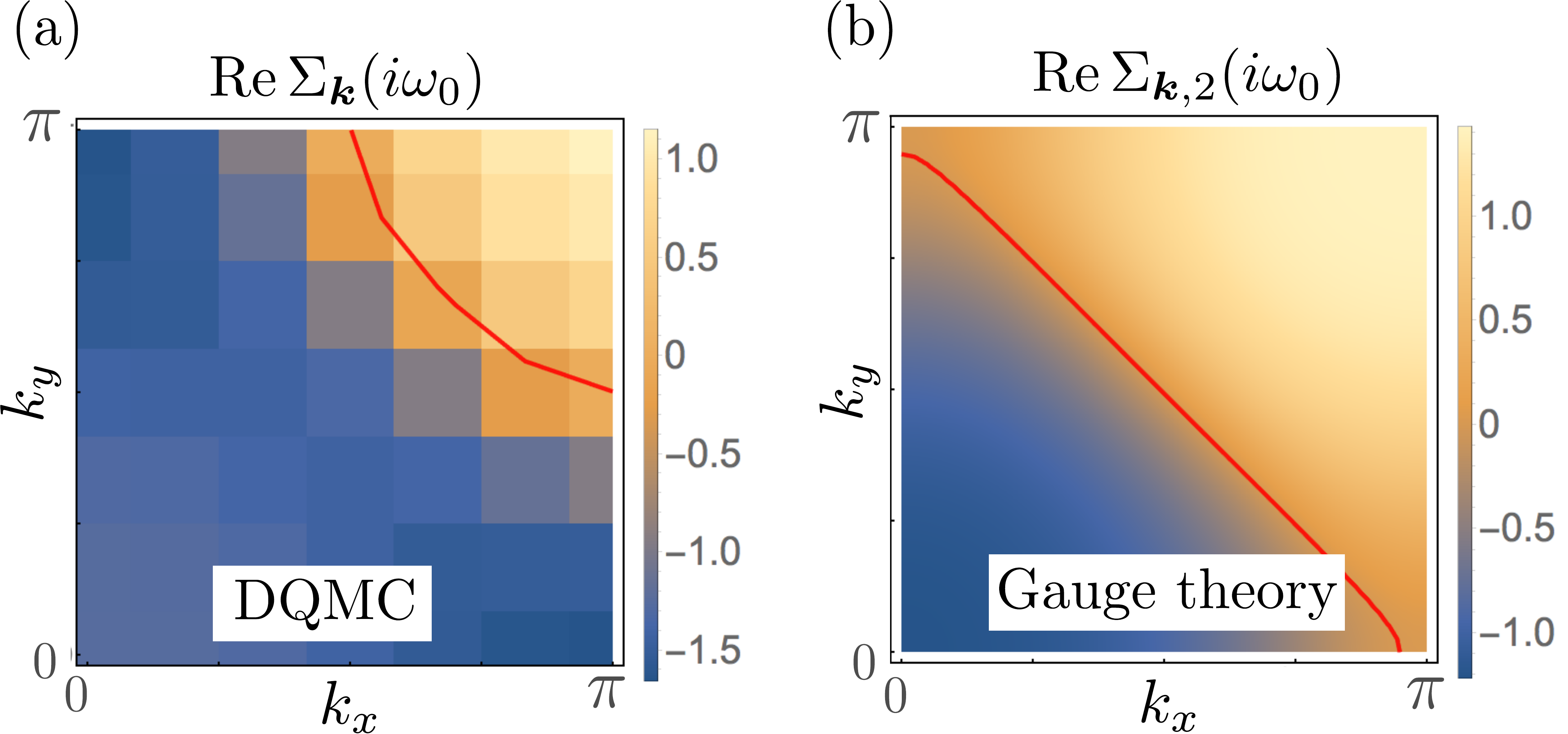}
\caption{The real part of the self energy (at the first Matsubara frequency) and its zeros indicated as red lines obtained (a) from DQMC  and (b) from the gauge theory. We use the same parameters as in the corresponding plot, Fig.~\eqtx{5}, of the main text.}
\label{CompareQMCRe}
\end{center}
\end{figure}

We first note that the DQMC result, shown in \figref{CompareQMCRe}(a), exhibits the sign changes from $(0,0)$ to $(\pi,\pi)$ that one expects from the form of the chargon self-energy in  Eq.~(\eqtx{10}).
However, a direct \change{quantitative} comparison to the predictions of the full SU(2) gauge theory is difficult as the real part of the self-energy (as opposed to the imaginary part) is crucially affected by the values of $Z_{i-j}$; It depends on the relation between the chargon dispersion, $\xi_{\vec{k}}$, and that of the electrons, $\epsilon_{\vec{k}}$. \change{This is the reason why we have mainly focused on the imaginary part in the remainder of the paper.}
 
To understand this issue better, let us distinguish two contributions to the self-energy,
\begin{equation}
\Sigma^\retarded_{\vec{k}} = \Sigma^\retarded_{\vec{k},1} + \Sigma^\retarded_{\vec{k},2}.
\end{equation}
Here $\Sigma^\retarded_{\vec{k},1} =\xi_{\vec{k}} - \epsilon_{\vec{k}}$ comprises all ``trivial'' renormalizations of the hopping parameters of the chargon dispersion (due to $Z_{i-j}\neq 1$). All other effects, that go beyond these band renormalizations, are contained in $\Sigma^\retarded_{\vec{k},2}$.  

If $Z_{i-j}$ are sufficiently small, $\Sigma^\retarded_{\vec{k}}$ is dominated by $\Sigma^\retarded_{\vec{k},1}$. In this limit, the sign changes of $\text{Re}\, \Sigma^\retarded_{\vec{k}}$ in the Brillouin zone are trivially determined by $-\epsilon_{\vec{k}}$ and exactly opposite to what we have in \figref{CompareQMCRe}(a) or expect from the chargon self-energy in Eq.~(\eqtx{10}).  
This is also what we find when taking the values of $Z_{i-j}$ from \equref{DeterminingZs} \change{for parameters in the physically relevant regime}; However, slightly increasing these factors, will lead to sign changes consistent with \figref{CompareQMCRe}(a).
These small differences are clearly beyond the accuracy to which we can determine $Z_{i-j}$ as we have already seen when comparing the gauge theory with DCA in the main text.

Instead of manually changing the values of $Z_{i-j}$, let us extract the ``nonrivial part'' $\Sigma^\retarded_{\vec{k},2}$ of the electron self-energy that comprises all effects beyond the aforementioned band renormalizations and does not crucially depend on the values of $Z_{i-j}$. Its real part is shown in \figref{CompareQMCRe}(b) and exhbits the same qualitative behavior as the real part of the self-energy obtained in DQMC.

\bibliography{NearZerosV2.bib}

\end{document}